\documentclass[11pt]{article}

\usepackage{verbatim}
\usepackage{amsmath}
\usepackage{amsfonts}
\usepackage{amssymb}
\usepackage{bbm}
\usepackage{latexsym}
\usepackage{lscape} 
\usepackage{epsfig}
\usepackage{color}
\usepackage{graphicx}
\usepackage[table]{xcolor}
\usepackage{enumerate}
\usepackage{hhline}
\usepackage{subfig}
\usepackage{multirow}
\usepackage{mathrsfs}
\usepackage{longtable}
\usepackage{hhline}

\usepackage[titletoc,title]{appendix}
\usepackage{amscd}
\usepackage{cite}

\DeclareMathSymbol{\mg}{\mathrel}{symbols}{"1D}

\newcommand{\ga}{\alpha}

\newcommand{\gd}{\delta}

\newcommand{\gk}{\kappa}

\newcommand{\cF}{{\cal F}}

\newcommand{\cN}{{\cal N}}
\newcommand{\cO}{{\cal O}}

\newcommand{\cR}{{\cal R}}

\newcommand{\cV}{{\cal V}}

\newcommand{\tr}{\text{tr}}

\newcommand{\ra}{\rightarrow}

\newcommand{\undr}[1]{{\underline{#1}}}

\newcommand{\beq}{\begin{equation}}
\newcommand{\eeq}{\end{equation}}
\newcommand{\barr}{\begin{array}}
\newcommand{\earr}{\end{array}}
\newcommand{\equ}[1]{\begin{gather} #1 \end{gather}}

\newcommand{\enums}[1]{\begin{enumerate} #1 \end{enumerate}}
\newcommand{\tabu}[2]{\begin{tabular}{#1} #2 \end{tabular}}
\newcommand{\arry}[2]{\begin{array}{#1} #2 \end{array}}

\newcommand{\non}{\nonumber}
\newcommand{\sfrac}[2]{\mbox{$\frac{#1}{#2}$}}
\newcounter{oldcounter}

\newcommand{\tga}{{\tilde \alpha}}

\newcommand{\Natr}{\mathbb{N}}
\newcommand{\Intr}{\mathbb{Z}}

\newcommand{\ba}[2]{\[\begin{array}{#2}\label{#1}}
\newcommand{\ea}{\end{array}\]}
\newcommand{\be}{\begin{equation}}
\newcommand{\ee}{\end{equation}}
\newcommand{\bea}{\begin{eqnarray}}
\newcommand{\eea}{\end{eqnarray}}

\newcommand{\E}[1]{\mathrm{E_{#1}}}
\newcommand{\U}[1]{\mathrm{U(#1)}}

\newcommand{\rep}[1]{\mathbf{#1}}
\newcommand{\crep}[1]{\overline{\rep{#1}}}
\newcommand{\brep}[1]{\overline{\rep{#1}}}

\newcommand{\sm}{{\,\mbox{-}}}

\usepackage{hyperref} 
\usepackage{cancel}

\begin{document}

\thispagestyle{empty}

\begin{flushright}
DESY-15-251, LMU-ASC 78/15
\\
\end{flushright}
\begin{center}
{\Large {\bf 
Line bundle embeddings for heterotic theories
} 
}
\\[0pt]

\bigskip
\bigskip 
{\large
{\bf{Stefan Groot Nibbelink}$^{a,}$\footnote{E-mail: Groot.Nibbelink@physik.uni-muenchen.de}},  
{\bf{Fabian Ruehle$^{b,}$}\footnote{E-mail: fabian.ruehle@desy.de}}
\bigskip}\\[0pt]
\vspace{0.23cm}
${}^a$ {\it 
Arnold Sommerfeld Center for Theoretical Physics,   \\ 
Ludwig-Maximilians-Universit\"at M\"unchen, 80333 M\"unchen, Germany
}
\\[1ex]
${}^b$ {\it 
Deutsches Elektronen-Synchrotron DESY, Notkestrasse 85, 22607 Hamburg, Germany
}
\\[1ex] 
\end{center}

\begin{abstract}
\noindent 
In heterotic string theories consistency requires the introduction of a non-trivial vector bundle. This bundle breaks the original ten-dimensional gauge groups $\text{E}_8\times\text{E}_8$ or $\text{SO}(32)$ for the supersymmetric heterotic string theories and $\text{SO}(16)\times\text{SO}(16)$ for the non-supersymmetric tachyon-free theory to smaller subgroups. A vast number of MSSM-like models have been constructed up to now, most of which describe the vector bundle as a sum of line bundles. However, there are several different ways of describing these line bundles and their embedding in the ten-dimensional gauge group. We recall and extend these different descriptions and explain how they can be translated into each other.
\end{abstract}

\newpage 
\setcounter{page}{1}
 \setcounter{footnote}{0}
\tableofcontents
\newpage 

\section{Introduction} 
\label{sc:Intro} 

There are three known consistent tachyon-free heterotic string theories: the two supersymmetric heterotic $\text{E}_8\times\text{E}_8$ and $\text{SO}(32)$ theories \cite{Gross:1984dd,Dixon:1985jw} and the non-supersymmetric $\text{SO}(16)\times\text{SO}(16)$ theory \cite{Dixon:1986iz,Dixon:1986jc,AlvarezGaume:1986jb}. Since these ten-dimensional gauge groups contain popular GUT groups, like SU(5) or SO(10), heterotic string theory is very well-suited for the study of string phenomenology and physics beyond the standard model.

Upon compactification on smooth Calabi--Yau manifolds, the Bianchi identities of the three-form field strength necessitate the introduction of a non-trivial vector bundle~\cite{Candelas:1985en}. The low energy gauge group $G$ that results from embedding this bundle in the ten-dimensional gauge group $\mathcal{G}$ is given by the commutant of $\mathcal{G}$ with the structure group $H$ of the bundle, $\mathcal{G}\rightarrow G\times H$. Besides the Bianchi identities the flux also has to satisfy the Donaldson-Uhlenbeck-Yau equations \cite{Donaldson:1985,Uhlenbeck:1986} to guarantee unbroken supersymmetry in the vacuum. On top of having to satisfy all of the above consistency conditions, the vector bundle influences directly many other phenomenological aspects such as the number of zero modes and consequently the number of families. Nevertheless quasi-realistic models were obtain in this fashion, see e.g.\ \cite{Bouchard:2005ag,Braun:2005ux}.

Most unfortunately there is at the moment no known mechanism that gives preference to a specific gauge bundle or compactification manifold. Consequently, the current state of the art is that bundles are chosen such that the resulting models resemble our observations as closely as possible. The construction of non-Abelian vector bundles on smooth Calabi-Yau manifolds has proven to be very involved, see e.g.\ \cite{Donagi:1999ez,Donagi:2000zs}. Instead, one can consider monad and Abelian line bundle backgrounds which are much simpler~\cite{Blumenhagen:2005zg,Blumenhagen:2006ux,Anderson:2007nc,Anderson:2012yf}. However, even in these cases, imposing all consistency conditions on the bundle simultaneously leads to a very complicated set of coupled Diophantic equations which can in general not be solved in reasonable amounts of time. 

In order to make nevertheless some progress, computer-aided searches are extremely crucial. The idea behind them is to scan over a huge amount of vector bundles and compactification spaces by randomly generating bundles such that they satisfy as many constraints as possible simultaneously by construction and then check whether the others are satisfied by chance. Despite being rare and relying on chance, huge amounts of quasi-realistic (MS)SM models have been constructed using line bundle backgrounds on both smooth orbifold resolutions~\cite{Honecker:2006qz,Nibbelink:2009sp,Blaszczyk:2010db} and on Complete Intersection Calabi-Yau manifolds (CICYs)~\cite{Anderson:2011ns,Anderson:2012yf,Anderson:2013xka,Nibbelink:2015ixa,Nibbelink:2015vha} even for compactifications of the non-supersymmetric heterotic string~\cite{Blaszczyk:2014qoa,Blaszczyk:2015zta}. 

As a next step one could then systematically analyze these models with regard to common features in order to see whether all string models constructed this way share a tendency towards certain properties. Since the amount of data is quite sizable, it is desirable to describe the bundles as efficiently as possible. Unfortunately the notion of efficiency is ambiguous, as different descriptions might be advantageous for different analyses. In addition, different authors use different conventions for describing the bundles, making the results hard to compare.

\subsubsection*{Outline and summary of the paper}

We study different descriptions of vector bundles as sum of line bundles. We explain how to define the embedding in the ten-dimensional gauge group and how to extract from them the data needed for further analyses. In addition, we compare them with regard to applicability and redundancy. Finally, we explain how the various descriptions can be translated into one another. The paper is organized as follows: In Section \ref{sc:CYs} we briefly review some properties of heterotic strings compactified on Calabi-Yau manifolds and set our notation. Sections \ref{sc:AbelianFluxes} to \ref{sc:NonAbelianMaxEmbedding} explain three possibilities of describing the bundle embedding. In Section \ref{sc:MatchingBundles} we discuss how the different descriptions can be related, before giving examples in Section \ref{sc:examples}. Appendix \ref{sc:GroupTh} contains a more detailed overview of group theory based on standard literature such as \cite{Slansky:1981yr}. In Appendix \ref{sc:IdUbundles} we repeat the matching procedure between line bundle vectors in two different bases as derived in~\cite{Blaszczyk:2015zta}.

\section{Smooth heterotic Calabi-Yau compactifications} 
\label{sc:CYs}

\begin{table}
\begin{center} 
\tabu{|c||c|c|}{
\hline 
{\bf Theory / group} &{\bf Properties} & {\bf Roots} \\ 
\hline
\hline 
Spin(32)/$\Intr_2$ & {N=1 SUSY} & $\big(\,\undr{\pm 1^2, 0^{14}}\,\big)$ 
\\[.5ex] \hline 
\multirow{2}{*}{E$_8$ $\times$ E$_8$} & \multirow{2}{*}{N=1 SUSY} &  
$\big(\,\undr{\pm 1^2, 0^{6}}\,\big)\big(\,0^8\,\big)$~,~$\big(\,\undr{-\sfrac12^{2k}, \sfrac12^{8-2k}}\,\big)\big(\,0^8\,\big)$~;  
\\ 
&& 
$\big(\,0^8\,\big)\big(\,\undr{\pm 1^2, 0^{6}}\,\big)$~,~$\big(\,0^8\,\big)\big(\,\undr{-\sfrac12^{2k}, \sfrac 12^{8-2k}}\,\big)$
\\[.5ex] \hline 
SO(16) $\times$ SO(16) &  \tabu{c}{Tachyon-free \\ non-SUSY} &  $\big(\,\undr{\pm 1^2, 0^{6}}\,\big)\big(\,0^8\,\big)$~;~$\big(\,0^8\,\big)\big(\,\undr{\pm 1^2, 0^{6}}\,\big)$
\\ \hline 
}
\end{center} 
\caption{
This table summarizes the properties of all tachyon-free ten-dimensional heterotic string theories and classifies them as being either supersymmetric or non-supersymmetric. The corresponding gauge group and the roots are indicated; the semicolons separate the different roots associated to the different gauge group factors. The underscore denotes all permutation of the entries.
\label{tb:10DHetGroups}}
\end{table}

\subsection{Ten-dimensional heterotic string theories}

To keep our discussion general so that it can be applied to either of the two supersymmetric heterotic string theories as well as to the non-supersymmetric one, we refer to the ten-dimensional gauge group as $\mathcal{G}$. For specific results of a given theory we will clearly specify which ten-dimensional theory is the starting point. 

In Table~\ref{tb:10DHetGroups} we list all known perturbative ten-dimensional heterotic string theories. We indicate whether they are supersymmetric or not. In addition, we give the roots associated to the non-Abelian generators. In order to have a universal description for all heterotic string theories, it is convenient to choose a Cartan subalgebra such that these roots can either be written as SO($2N$)-adjoint roots $\big(\,\undr{\pm 1^2,0^{N-2}}\,\big)$ or SO($2N$)-spinor weights $\big(\,\undr{-\sfrac 12^{2k}, \sfrac12^{N-2k}}\,\big)$, $k \in \Natr$. Further details on this notation are collected in Appendix~\ref{sc:Lattices}

The massless spectra of the two supersymmetric heterotic string theories are well-known and easily summarized: their supergravity sector contains the graviton, B-field, dilaton and their superpartners, the gravitino and the dilatino. Their super-Yang-Mills sectors contain gauge fields and their gauginos in the adjoint representation of the respective gauge group. The non-supersymmetric SO(16)$\times$SO(16) theory has the same universal gravitational and gauge sector, i.e.\ graviton, B-field, dilaton and the gauge fields in the appropriate adjoint representations, but none of their superpartners! Nevertheless this tachyon-free non-supersymmetric heterotic string theory has a fermionic spectrum in ten dimensions given in Table~\ref{tb:FermionsSO16xSO16} which is free of anomalies. 

\begin{table} 
\[
\arry{|c||c|c|c|}{
\hline 
& \multicolumn{3}{c|}{\text{\bf Fermionic states of the non-SUSY SO(16)$\times$SO(16) theory}} 
\\ \hline\hline 
  \text{\bf Repr.}  & (\rep{128}; \rep{1})_+  & (\rep{1}; \rep{128})_+ & (\rep{16}; \rep{16})_- 
\\ \hline  
 \text{\bf Weights}  &
 \big(\,\undr{-\sfrac12^{2k}, \sfrac12^{8-2k}}\,\big)\big(\,0^8\,\big) 
& 
 \big(\,0^8\,\big)\big(\,\undr{-\sfrac12^{2k}, \sfrac12^{8-2k}}\,\big)
& 
 \big(\,\undr{\pm1,0^7}\,\big)\big(\,\undr{\pm1,0^7}\,\big)
\\ \hline 
}
\non 
\]
\caption{ \label{tb:FermionsSO16xSO16}  
This table lists the non-Abelian representations and the corresponding weights $p$ for the non-supersymmetric SO(16)$\times$SO(16) theory. The $\pm$-subscript gives the ten-dimensional chirality of these states.}
\end{table}

\subsection{General topological characterization of Calabi-Yau manifolds}

A complex three-dimensional K\"ahler manifold $X$ with vanishing first Chern class, $c_1=0$, is called a Calabi-Yau threefold. Such manifolds are crudely characterized by two independent Hodge numbers $h_{11}$ and $h_{21}$ counting the number of closed but not exact (1,1)- and (2,1)-forms, respectively. The Hodge number $h_{21}$ counts the number of independent complex structure deformations that $X$ admits. The number of linearly independent divisors $D_i$, complex codimension one hypersurfaces of $X$, is counted by $i = 1,\ldots,h_{11}$. (Since we will have many different types of indices in this paper, we have collected our conventions in Table~\ref{tb:IndexConventions}.) Their dual curves $C_i$ define complex codimension two subspaces. It is often convenient to choose a minimal integral basis for these divisors and curves such that 
\equ{ \label{IntegralBasis} 
\int_{C_i} D_j = \int_{D_i} C_j = \delta_{ij}\,. 
}
The integrals here are defined over the Poincar\'e-dual two- and four-forms, respectively. For notational simplicity we use the same symbols to refer to either a divisor as a hypersurfaces or to the corresponding two-form, since it becomes clear from the context which one is meant. The intersection of two linear inequivalent divisors leads to a curve $C_{ij} = D_i D_j$. The triple intersections of divisors are called intersection numbers and are defined as 
\equ{ \label{Intersection}
\kappa_{ijk}(X) = \int_X D_i D_j D_k\,.
}
Since in this formula the divisors denote the corresponding two-forms, it can also be used to define self-intersection numbers.

Further characterizations of a Calabi-Yau manifold are provided via its Chern classes. The first Chern class vanishes for a Calabi-Yau manifold by definition. For the second and third Chern classes, $c_2$ and $c_3$,  we can define the topological numbers
\equ{ \label{Chern}
c_{2i}(X)  = \int_{D_i} c_2 = \int_X D_i c_2 \,, 
\quad 
c_3(X) = \int_X c_3\,. 
}

The K\"ahler form $J$ is a harmonic two-form
\equ{ 
J = a_i\, D_i\,, 
}
and can therefore be expanded in a basis of linearly independent divisors; the expansion coefficients $a_i$ are called K\"ahler moduli. The K\"ahler form can be used to determine the volumes of curves $C$, divisors $D$ and the Calabi-Yau $X$ itself via  
\equ{
\text{Vol}(C) = \int_C J\,, 
\quad 
\text{Vol}(D) = \sfrac 12 \int_D J^2\,,
\quad 
\text{Vol}(X) = \sfrac 16 \int_ X  J^3\,,
}
respectively. 

There are various types of Calabi-Yau manifolds that are frequently considered in the literature~\cite{CYweb}: CICYs provide a large class of well-studied geometries~\cite{Candelas:1987kf,Braun:2010vc}. Another popular list, originally due to Kreuzer and Skarke~\cite{Kreuzer:2000xy} can now be searched for various criteria electronically~\cite{Altman:2014bfa}. In addition, orbifold resolutions~\cite{Lust:2006zh,Nibbelink:2007pn,Nibbelink:2009sp,Blaszczyk:2010db,Nibbelink:2012de} lead to a collection of Calabi-Yau spaces with relatively large $h_{11}$.

\begin{table}
\begin{center}
\tabu{|l|p{12cm}|}{
\hline 
\textbf{Indices} & \textbf{Description} \\ \hline\hline 
$i,j,\ldots = 1,\ldots, h_{11}$ & label the $h_{11}$ divisors of a Calabi-Yau space 
\\\hline 
$I,J,\ldots = 1,\ldots, 16$ & label the Cartan generators of the ten-dimensional gauge group
\\\hline 
$a,b,\ldots = 1,\ldots, r$ & label the U(1) factors of a rank $r$ line bundle
\\\hline 
$\tilde{a}, \tilde{b}, \ldots = 1,\ldots, r+1$ & 
label the nodes of the extended Dynkin diagram of a rank $r$ algebra; for $\text{A}_r$ they can be identified with the vector representation indices
\\\hline 
}
\end{center}
\caption{\label{tb:IndexConventions} 
This table summarizes our index conventions used throughout this paper. Moreover, unless otherwise stated, summation over repeated indices (using the standard Euclidean metric) is implied.}
\end{table}

\subsection{Line bundles on Calabi-Yau spaces}

A line bundle on a Calabi-Yau manifold $X$ is denoted by
\equ{ 
\cV = \cO_X(q_1, \ldots, q_{h_{11}})\,. 
}
In the minimal integral divisor basis we have 
\equ{ 
\label{eq:DefLs}
\int_{C_i} c_1(\cV) = q_i\,, 
\qquad c_1(\cV) = q_i\, D_i\,, 
}
i.e.\ the numbers $q_i$, $i=1,\ldots, h_{11}$, specify the first Chern classes integrated over an appropriate basis of curves. This can be extended to a direct sum of line bundles,  
\equ{
\label{eq:BundleDescription}
\mathcal{V} = \bigoplus_{a=1}^r
 \mathcal{O}_X\big(q^{a}_1,\ldots, q^{a}_{h_{11}}\big)\,, 
}
with  $a=1,\ldots,r$, which leads to a rank $r$ bundle. The integers $q^a_i$ define a matrix $q = (q^a_i)$ that characterizes the U(1)$^r$ line bundle for a specific choice of embedding. (The index conventions can be found in Table~\ref{tb:IndexConventions}.)

\section{Abelian gauge fluxes} 
\label{sc:AbelianFluxes}

\subsection{Gauge background and consistency conditions}

We can describe an Abelian gauge background embedded in the ten-dimensional gauge group $\mathcal{G}$ by expanding its field strength, 
\equ{  \label{LineBundleFlux} 
 \frac{\mathcal{F}}{2\pi} = D_i\, H_i\,. 
 }
Since the Hermitian Yang-Mills equations demand that $\cF$ is a (1,1)-form, we can expand it in the (1,1)-forms dual to the divisors $D_i$. Furthermore, as the background is assumed to be Abelian we can decompose the algebra-valued coefficients~ 
\equ{ \label{eq:LineBundleEmbeddinginG} 
 H_i = V_i^I\, H_I\,, 
}
with $I = 1,\ldots, 16$, in terms of a Cartan subalgebra of the ten-dimensional gauge group $\mathcal{G}$ generated by $H_I$ \cite{Strominger:1986uh,Nibbelink:2007rd,Nibbelink:2007pn}. More specifically, given the representation of the roots listed in Table~\ref{tb:10DHetGroups} for the E$_8\times$E$_8$ theory, the $H_I$ denote the Cartan generators of its maximal SO(16)$\times$SO(16) subgroup. They are chosen such that
\begin{align} \label{eq:HINormalization}
 \tr\, H^I H^J=\delta^{IJ}\,.
\end{align}
The trace here is in the adjoint representation of the respective ten-dimensional gauge group but normalized as if they were SU-generators. 

Consequently, the Abelian gauge background of a heterotic string theory can be characterized by $h_{11}$ sixteen-component vectors $V_i = (V_i^I)$, which are often referred to as line bundle vectors. In order to refer to the two factors in the E$_8\times$E$_8$ or SO(16)$\times$SO(16) theories separately we employ the notation $V_i = (V^{\prime}_i, V^{\prime\prime}_i)$ to decompose the line bundle vectors $V_i$ into observable and hidden parts, respectively. 
 
So far the line bundle vectors $V_i$ seem to be totally arbitrary vectors. However, for consistency of the line bundle background they have to satisfy various properties:

\subsubsection*{Flux quantization}

First of all the flux background~\eqref{LineBundleFlux} has to be integrally quantized on all states of the respective theories. This means that 
\equ{ \label{LineBundleFluxQuantization} 
\int_C  \frac{\mathcal{F}}{2\pi} (p) \in \Intr\,, 
}
for any curve $C$ evaluated on any state of the theory with weight $p = (p_I)$. In the minimal integral basis of divisors and curves this reduces to  
\equ{ \label{FluxQuantizationIntBasis} 
H_i(p) = V_i \cdot p = V_i^I\, p_I  \in \Intr\,. 
}
The appropriate lattices for the three theories are indicated in Table~\ref{tb:FluxLattices} where we use the four types of lattices given in Table~\ref{tb:Lattices} of Appendix~\ref{sc:Lattices}. 

If the gauge flux is integral on all massless states in ten dimensions it is an admissible background for the low energy limits of the various heterotic string theories. However, it might not have a lift to string theory. As a necessary condition to guarantee a full string lift, the gauge flux has to be integrally quantized on any state in the string spectrum. This difference is subtle yet important: To ensure integral quantization on all massless states of the Spin(32)$/\Intr_2$ theory it is sufficient that $V_i \in \mathbf{R}_{16} \oplus \mathbf{V}_{16} \oplus \mathbf{S}_{16} \oplus \mathbf{C}_{16}$ are vectors with either all integral or all half-integral entries whose sum is an even integer.

\begin{table}
\begin{center} 
\tabu{|c||c|}{
\hline 
{\bf Theory / group} &{\bf Lattice} 
\\ \hline\hline 
Spin(32)/$\Intr_2$ & $\mathbf{R}_{16} \oplus \mathbf{S}_{16}$ 
\\[1ex]   
E$_8$ $\times$ E$_8$ & $\big(\mathbf{R}_{8} \oplus \mathbf{S}_{8} \big)$ $\otimes$ $\big(\mathbf{R}_{8} \oplus \mathbf{S}_{8} \big)$  
\\[1ex]  
SO(16) $\times$ SO(16) &  $\big(\mathbf{R}_8\otimes\mathbf{R}_8\big)$ $\oplus$ $\big( \mathbf{S}_8 \otimes \mathbf{S}_8\big)$ 
\\ \hline 
}
\end{center} 
\caption{
This table lists the lattices which the line bundle vectors have to embed into in the minimal integral basis. 
\label{tb:FluxLattices}}
\end{table}

\subsubsection*{Bianchi identities}

Not all integrally quantized line bundle backgrounds are well-defined. Consistency of the heterotic compactification requires that the Bianchi identities of the 3-form field strength $H$,
\equ{ \label{BInonInt}
\text{ch}_2(\cF) - \text{ch}_2(\cR) = [W]\,,
}
are satisfied where $[W]$ is a curve class (respectively its dual 4-form class) and $\text{ch}_2(\cF)$ and $\text{ch}_2(\cR)$ denote the second Chern characters of the vector and tangent bundle. Using \eqref{eq:HINormalization}, the second Chern character is simply given by
\begin{align}
 \label{eq:secondChernCharacterBundle}
 \text{ch}_2(\cF)=-\sfrac12\, V_i \cdot V_j\, D_i D_j\,,
\end{align}
where the $D_i$ are two-forms and the dot denotes the standard scalar product. The Bianchi identities ensure that the resulting four-dimensional theory is free of irreducible anomalies. The reducible anomalies are cancelled via the Green-Schwarz mechanism. 

When $[W]$ is non-trivial, there are NS5 branes wrapping the corresponding curve $W$. Their wrapping numbers can be computed as 
\equ{ 
[W] = N_i \, C_i\,, 
\qquad 
N_i = \int_{D_i} W
}
in the minimal integral basis. In order that the NS5 branes preserve the same supersymmetries as the perturbative part of the supersymmetric heterotic string theories, the NS5-branes need to wrap effective curves and hence their charges $N_i$ need to be non-negative. 

The integrated version of the Bianchi identities~\eqref{BInonInt} impose conditions on the line bundle vectors, 
\equ{  \label{BIs}
N_i =  \kappa_{ijk}\, V_j \cdot V_k + 2\, c_{2i}\,,
}
for all divisors $D_i$\,, $i=1,\ldots, h_{11}$. For the non-supersymmetric SO(16)$\times$SO(16) theory, the condition to preserve supersymmetry is obsolete. Nevertheless, in order to ensure that the (anti-)NS5 branes do not introduce any tachyons, we require their absence, $N_i=0$, to be on the safe side.

\subsubsection*{DUY constraints}

The final conditions an Abelian gauge flux has to satisfy are the DUY equations
\equ{ 
0 = \frac 12 \int_X J^2\, \frac{\cF}{2\pi} = \text{Vol}(D_i) \, H_i\,, 
}
with all divisor volumes $\text{Vol}(D_i) > 0$. Even though this just seems to be a set of linear equations on these volumes, they can often be a very constraining conditions on the line bundle vectors themselves. 

The DUY equations quoted above are the ones at tree level; the one-loop corrections are known in the supersymmetric case, see e.g.\ \cite{Blumenhagen:2005ga,Blumenhagen:2005pm}. For the non-supersymmetric case, we follow ref.~\cite{Blaszczyk:2014qoa,Blaszczyk:2015zta} to exploit the fact that on supersymmetric backgrounds at tree-level and leading order in $\ga'$, equations of motion for the bosonic fields of the supersymmetric and the non-symmetric theories are identical (up to the gauge groups), which means that supersymmetry preserving backgrounds are also backgrounds of the non-supersymmetric SO(16)$\times$SO(16) string theory. At the one-loop level one expects that non-supersymmetric effects become apparent. In particular, one expects that corrections to the DUY equations cannot be calculated along the lines of refs.\ \cite{Blumenhagen:2005ga,Blumenhagen:2005pm}, since these computations rely on supersymmetry.

\subsection{Unbroken gauge group and spectrum}

The unbroken gauge group $G$ in four dimensions is given by the commutant of the structure group $H$ of the bundle with the ten-dimensional gauge group $\mathcal{G}$: $\mathcal{G}\supset G\times H$.
The non-Abelian part of $G$ can be computed by determining all roots $p$ of the ten-dimensional gauge fields, given in Table~\ref{tb:10DHetGroups}, that are perpendicular to all line bundle vectors, i.e.\ 
\equ{ \label{GaugeGroup}
H_i(p) = V_i \cdot p \stackrel{!}{=} 0\,, 
}
for all $i = 1,\ldots,h_{11}$. Since we consider vector bundles that are sums of Abelian bundles, $H$ is Abelian and commutes. However, the U(1)s in $H$ are generically massive due to one-loop effects~\cite{Blumenhagen:2005ga,GrootNibbelink:2007ew,Nibbelink:2009sp}; nevertheless they remain as global selection rules.

\subsubsection*{Chiral spectrum} 

To compute the chiral part of the fermionic spectrum we can make use of the multiplicity operator 
\equ{ \label{MultiOp}
\mathcal{N}  = \sfrac16\, \kappa_{ijk}\, H_i H_j H_k + \sfrac1{12}\, c_{2i}\,  H_i\,. 
}
This operator can be evaluated on every weight $p$ of the given fermionic representations.  A left-chiral fermion in four dimensions has $\mathcal{N}(p) >0$ while its right-chiral CPT partner with weight $-p$ has the same multiplicity with opposite sign. 

The multiplicity operator can also be used to compute part of the bosonic spectrum. For the supersymmetric heterotic string theories this just determines the bosonic superpartners of the massless fermions determined by the fermionic multiplicity operator. For the non-supersymmetric \linebreak SO(16)$\times$SO(16) theory we can evaluate the multiplicity operator on the roots of the ten-dimensional gauge fields to determine the number of the various irreducible representations of massless complex scalars.

\subsubsection*{Redundancies} 

The description of line bundles presented in this section unambiguously characterizes how the line bundle background is embedded in the ten-dimensional gauge group $\mathcal{G}$. However, it suffers from a large number of redundancies. They arise from listing explicitly the embedding of the bundle with respect to the 16 Cartan generators of $\mathcal{G}$. If the vector bundle has rank $r<16$ and is given by a sum of line bundles, the bundle has only $r$ independent Cartan directions, which are a linear combination of the 16 Cartan direction in $\mathcal{G}$. Furthermore, the bundle is given with respect to an arbitrary choice of the 16 Cartan generators in the ten-dimensional gauge algebra. This by itself does not introduce any redundancy, but makes the notation dependent on the embedding. Another source for potential redundancy is that seemingly completely different line bundles could potentially be related by Weyl reflections, which are symmetry operations on the root lattice. The order of the Weyl group can be very big; of the order of $7\times 10^{8}$ for E$_8$ and $7\times 10^{17}$ for SO(32), respectively. When constructing models, great care has to be taken in order to not overcount the number of constructed models.

\section{Line bundles with Levi embedding} 
\label{sc:LeviEmbedding} 

Given that the transition functions of line bundles are U(1) phases, there is a natural way to embed the rank $r$ bundle~\eqref{eq:BundleDescription} in the ten-dimensional gauge group $\mathcal{G}$ of a heterotic string theory.  Since U(1) gauge backgrounds are rank-preserving (up to the fact that U(1) factors can acquire a mass via the St\"uckelberg mechanism upon Green-Schwarz anomaly cancellation)
we have 
\equ{
\mathcal{G} \supset G \times U(1)^r\,, 
\quad 
\text{with} 
\quad
\text{rank}(G)+r = 16\,. 
}
This embedding can be defined as the Levi embedding (see Appendix~\ref{sc:Branchings} for details): We label the $r$ nodes that are removed from the ordinary Dynkin diagram of $\mathcal{G}$ by $a$. Removed nodes are turned into U(1) factors. Let us denote the Cartan generators that correspond to the simple roots $\ga_a$ which are dropped by removing the nodes by $h_a$, such that
\equ{ \label{LeviCartanBasis}
h_a(\ga_b) = \gd_{ab}\,, 
\qquad 
h_a =  (A^{-1})_{ab}\, \alpha_b^I \, H_I\,, 
}
for all simple roots $\ga_b$ of $\mathcal{G}$. The matrix $A_{ab}$ is the Cartan matrix of $\mathcal{G}$. In particular, using~\eqref{eq:HINormalization} and~\eqref{LeviCartanBasis} as well as the definition of the Cartan matrix \eqref{CartanMatrix}, one finds
\begin{align}
 \tr\, h_a h_b=(A^{-1})_{ab}\,.
\end{align}
To make this choice of basis manifest we label the Chern classes in \eqref{eq:DefLs} that define $\mathcal{V}$ by $\ell_i^a$. The background gauge field strength in this embedding is given by 
\equ{
\label{eq:BundleFieldStrength}
\frac {\cF}{2\pi} =  D_i\, h_i\,, 
\qquad 
h_i = \ell_i^a\, h_a\,. 
}

\subsubsection*{Flux quantization}

In any description flux quantization is equivalent to~\eqref{eq:BundleFieldStrength} taking integral values when integrated over any curve $C$ and evaluated on any state of the heterotic string theory in question. From Tables~\ref{tb:10DHetGroups} and~\ref{tb:FermionsSO16xSO16} we hence find that
\equ{ \label{Quantization}
h_i(p) = \ell_i^a \, p_a \in \Intr\,, 
}
for all $i=1,\ldots, h_{11}$ and all weights $p$ associated with these states: $p_a = h_a(p) = (A^{-1})_{ab}\, \ga_b\cdot p$. This shows that for the E$_8\times$E$_8$ theory flux quantization is equivalent to demanding that all $\ell_i^a\in\Intr$. The simple roots of each E$_8$ span the E$_8$ root lattice $\mathbf{R}_8\oplus \mathbf{S}_8$. Since this root lattice is self-dual, the inner product of any to vectors from this lattice is integral. 

For the SO(32) theory we note that any state $p$ in the lattice given in Table~\ref{tb:FluxLattices} can be represented as 
\equ{
p \in \mathbf{R}_{16}\,,
\quad\text{or}\quad
p \in \sfrac 12\, e_{16} +  \mathbf{R}_{16}\,,
}
where $e_{d}=(1,\ldots, 1)$ is a vector with $d$ ones. Moreover, any vector in $\mathbf{R}_{16}$ can be spanned by the simple roots of SO(32). Evaluating~\eqref{Quantization} on any SO(32) simple root $\ga_c$ and on $\sfrac 12\, e_{16}$ leads to the conditions 
\equ{ 
\ell_i^c = \ell_i^a\,\, (A^{-1})_{ab}\, \ga_b\cdot \ga_c \in \Intr\,,
\qquad 
\ell_i^{a} \, (A^{-1})_{a\,16} = \ell_i^a\, (A^{-1})_{ab}\, \ga_b \cdot \sfrac 12\, e_{16} \in \Intr\,,
}
using that $\ga_b \cdot \sfrac 12\, e_{16} = \gd_{b\,16}$.

Finally, for the SO(16)$\times$SO(16) theory the charges $p$ of the states are from the lattices
\equ{
p \in \big( \mathbf{R}_8 \oplus \mathbf{S}_8 \big) \,\otimes\, 
 \big( \mathbf{R}_8 \oplus \mathbf{S}_8 \big)
\quad\text{or}\quad
\big( \mathbf{V}_8 \oplus \mathbf{C}_8 \big) \,\otimes\, 
 \big( \mathbf{V}_8 \oplus \mathbf{C}_8 \big)\,, 
}
see e.g.~\cite{Blaszczyk:2015zta}. Notice that 
\(
\big( \mathbf{V}_8 \oplus \mathbf{C}_8 \big) \,\otimes\, 
 \big( \mathbf{V}_8 \oplus \mathbf{C}_8 \big)
= (1,0^7)(1,0^7) + 
\big( \mathbf{R}_8 \oplus \mathbf{S}_8 \big) \,\otimes\, 
 \big( \mathbf{R}_8 \oplus \mathbf{S}_8 \big)\,,
\)
hence in this case we find for the first SO(16) factor: 
\equ{
\ell_i^{\prime a} 
\in \Intr\,,
\qquad 
\ell_i^{\prime a}\, (A^{\prime-1})_{a8}  \in \Intr\,, 
\qquad 
(\ell_i^{\prime a} +
\ell_i^{\prime\prime a})\, (A^{\prime-1})_{a1} \in \Intr\,, 
}
and similar for the second factor, i.e.\ replace everywhere ${}^\prime$ by ${}^{\prime\prime}$. Here we have used that $\ga^{\prime}_b\cdot (1,0^7) = \gd_{b1}$ and that the Cartan matrices for both factors are that of SO(16) and hence equal.

\subsubsection*{Four-dimensional spectrum} 

To determine the massless chiral spectrum, we first decompose the adjoint of the ten-dimensional gauge group
\equ{ 
\mathbf{ad}(\mathcal{G}) = \bigoplus_x \mathbf{R}_{x;q_x} 
}
into irreducible representations of $G$ with $r$ U(1) charges $q_x = (q_{x;1},\ldots,q_{x;r})$ computed via $q_{x;a} = h_a(p)$ where $p$ is a weight vector corresponding to $\mathbf{R}$. (For the fermions in the non-supersymmetric SO(16)$\times$SO(16) theory we can follow the same procedure for the representations given in Table~\ref{tb:FermionsSO16xSO16}.)  A convenient way to describe this in detail is provided by the projection matrix reviewed in Appendix~\ref{sc:Branchings} which relates the roots of the ten-dimensional gauge group to the roots and U(1) generators of the four-dimensional gauge group.

The multiplicities of the irreducible representation are then determined by their index 
\equ{ 
\cN(\mathbf{R}_x) = 
\sfrac 16\, \gk_{ijk}\, h_i h_j h_k + \sfrac 1{12}\, c_2\, h_i\,
}
which can be evaluated on any weight $p \in \mathbf{R}_x$ in the representation $\mathbf{R}_x$. 

\subsubsection*{Redundancies and applicability}
This description does not contain any redundancies and is always applicable. In order to fully specify the vector bundle we need to specify its embedding in $\mathcal{G}$, which we do e.g.\ by listing which simple roots get broken or by specifying the non-Abelian gauge group $G$. This fixes the embedding of $H$ into the Cartan subalgebra of $G$ via the condition \eqref{LeviCartanBasis}. By giving the first Chern class of the line bundles in this embedding on all $h_{11}$ divisors we completely fix the bundle.

\section{Line bundles with maximal non-Abelian enhancement}
\label{sc:NonAbelianMaxEmbedding} 

In certain cases one can consider the possibility that the line bundle background can -- at least group-theoretically -- be enhanced to a semi-simple group $\widetilde{H}$ such that $G \times \widetilde{H}$ is a maximal subgroup of the ten-dimensional gauge group $\mathcal{G}$. In this case we can specify the embedding of the line bundles in $\mathcal{G}$ in a two-step procedure
\equ{ 
\mathcal{G} \supset G \times \widetilde{H}\,, 
\qquad 
\widetilde{H} \supset \text{U}(1)^r\,. 
}
This has the advantage that the embedding of maximal subgroups in the group $\mathcal{G}$ is unique. Moreover, also the embedding of U$(1)^r$ into $\widetilde{H}$ is restricted quite a lot. The first step, i.e.\ the embedding of the semi-simple factors in $\mathcal{G}$, can be described using the projection matrix method as reviewed in Appendix~\ref{sc:Branchings}.

\subsubsection*{S(U(1)\texorpdfstring{$^{\boldsymbol{r}}$}{\^{}r}) $\subset$ SU($\boldsymbol{r}$) line bundle embeddings in $\boldsymbol{\mathcal{G}}=\text{E}_{\boldsymbol{8}}$}

\begin{table}[t]
\begin{center} 
\tabu{c}{
 \begin{tabular}{|c|l|l|}
 \hline
 \multicolumn{3}{|c|}{\bf Branching $\text{E}_8 \supset \text{E}_{8-r}\times \text{A}_r$} 
 \\ 
{\bf Algebra }					& {\bf Irreps} 				& $\cV \subset \text{A}_r$  				\\
 \hline
 \hline
 \multirow{3}{*}{$\text{E}_{7}\times A_1$} 	& $(\rep{133},\rep{1})$	 		& $\mathcal{O}$				\\
						& $(\rep{1},\rep{3})$ 			& $\mathcal{V}\otimes\mathcal{V}^*$	\\
						& $(\rep{56},\rep{2})$			& $\mathcal{V}$				\\
 \hline
 \multirow{3}{*}{$\text{E}_{6}\times \text{A}_2$} 	& $(\rep{78},\rep{1})$ 			& $\mathcal{O}$				\\
						& $(\rep{1},\rep{8})$ 			& $\mathcal{V}\otimes\mathcal{V}^*$	\\
						& $(\rep{27},\rep{3})$	& $\mathcal{V}$				\\
 \hline  
  	
						& $(\rep{45},\rep{1})$ 			& $\mathcal{O}$				\\
$\text{E}_{5}\times \text{A}_3$			& $(\rep{1},\rep{15})$ 			& $\mathcal{V}\otimes\mathcal{V}^*$	\\
($\text{E}_5 = \text{D}_5$)			& $(\rep{16},\rep{4})$		& $\mathcal{V}$				\\
						& $(\rep{10},\rep{6})$ 			& $\bigwedge^2\mathcal{V}$		\\
 \hline  
 \end{tabular}
\! 
  \begin{tabular}{|c|l|l|}
 \hline
 \multicolumn{3}{|c|}{\bf Branching $\text{E}_8 \supset \text{E}_{8-r}\times \text{A}_r$} 
 \\ 
{\bf Algebra }					& {\bf Irreps} 				& $\cV \subset \text{A}_r$  				\\
 \hline
 \hline
						& $(\rep{24},\rep{1})$ 			& $\mathcal{O}$				\\
 $\text{E}_{4}\times \text{A}_4$		& $(\rep{1}, \rep{24})$ 		& $\mathcal{V}\otimes\mathcal{V}^*$	\\
 ($\text{E}_4 = \text{A}_4$)			& $(\rep{10},\rep{5})$			& $\mathcal{V}$				\\
						& $(\brep{5},\rep{10})$ 		& $\bigwedge^2\mathcal{V}$		\\
 \hline  
						& $(\rep{3},\rep{1},\rep{1})$ 		& $\mathcal{O}$				\\
						& $(\rep{1},\rep{8}, \rep{1})$ 		& $\mathcal{O}$				\\
$\text{E}_{3}\times \text{A}_5$			& $(\rep{1},\rep{1}, \rep{35})$		& $\mathcal{V}\otimes\mathcal{V}^*$	\\
($\text{E}_3 = \text{A}_1\times\text{A}_2$)	& $(\rep{2},\crep{3}, \rep{6})$	& $\mathcal{V}$				\\
						& $(\rep{1},\crep{3}, \rep{15})$	& $\bigwedge^2\mathcal{V}$		\\
						& $(\rep{2},\rep{1}, \rep{20})$	& $\bigwedge^3\mathcal{V}$		\\
 \hline  
 \end{tabular}
}
\end{center} 
\caption{  \label{tab:extendedBranchingE8}
 Branching of the adjoint of E$_8$ into irreducible representations of $\text{E}_{8-k}\times \text{A}_k$. If both an irreducible representation and its conjugate appear in the branching we only list it once. We assume that the bundle structure group is SU($r+1$). 
}
\end{table}

\begin{table}[t]
\begin{center} 
 \begin{tabular}{|lc|l|}
 \hline
 \multicolumn{3}{|c|}{\bf Branching $\text{D}_{\boldsymbol N} \supset \text{D}_{\boldsymbol n} \times \text{D}_{\boldsymbol r}$, $\boldsymbol{N}=\boldsymbol{n}+\boldsymbol{r}$}
 \\
   \multicolumn{2}{|c|}{\bf $\text{D}_{\boldsymbol N}$ irrep }					& {\bf branched irreps}	\\
 \hline
  \multicolumn{2}{|c|}{$\boldsymbol{2N}$} & $(\boldsymbol{2n},\rep{1})+(\rep{1},\boldsymbol{2r})$
  \\
   \multicolumn{2}{|c|}{\bf ad$_{\text{D}_{\boldsymbol{N}}}$} & \bf $($ad$_{\text{D}_{\boldsymbol{n}}},\rep{1})+(\rep{1},$ad$_{\text{D}_{\boldsymbol{r}}})+(\rep{2n},\rep{2r})$\\
   \multicolumn{2}{|c|}{$\boldsymbol{2^{N-1}}$} & $(\boldsymbol{2^{n-1}_+},\boldsymbol{2^{r-1}_+})+(\boldsymbol{2^{n-1}_-},\boldsymbol{2^{r-1}_-})$\\
 \hline
 \multicolumn{3}{|c|}{\bf Branching $\text{D}_{\boldsymbol N} \supset \text{D}_{\boldsymbol n} \times \text{A}_{\boldsymbol{r-1}}\times \text{U}_1$, $\boldsymbol{N}=\boldsymbol{n}+\boldsymbol{r}$}\\
   \multicolumn{2}{|c|}{\bf $\text{D}_{\boldsymbol N}$ irrep }					& {\bf branched irreps}	\\
 \hline
   \multicolumn{2}{|c|}{$\boldsymbol{2N}$} & $(\boldsymbol{2n},\rep{1})_\rep{0}+(\rep{1},\boldsymbol{r})_\rep{1}+(\rep{1},\boldsymbol{\overline{r}})_\rep{-1}$
   \\
   \multicolumn{2}{|c|}{\bf ad$_{\text{D}_{\boldsymbol{N}}}$} & \bf $($ad$_{\text{D}_{\boldsymbol{n}}},\rep{1})_0\!+\!(\rep{1},$ad$_{\text{A}_{\boldsymbol{r-1}}})_0\!+\!(\rep{1},\rep{1})_0\!+\!(\rep{2n},\rep{r})_\rep{1}\!+\!(\rep{2n},\brep{r})_\rep{-1}\!+\!(\rep{1},\rep{\frac{r(r-1)}{2}})_\rep{2}\!+\!(\rep{1},\brep{\frac{r(r-1)}{2}})_\rep{-2}$
   \\
 \hline
 \multirow{4}{*}{\tabu{c}{$\boldsymbol{128};$\\[1ex]  ~~$\boldsymbol{r =}$}} \!\!\!\!& \!\!$2$ & $(\rep{32}_+,\rep{1})_\rep{1}+(\rep{32}_+,\rep{1})_\rep{-1}+(\rep{32}_-,\rep{2})_\rep{0}$\\
 & \!\!$3$ & $(\rep{16}_+,\rep{3})_\rep{-\frac12}+(\rep{16}_-,\brep{3})_\rep{\frac12}+(\rep{16}_+,\rep{1})_\rep{\frac32}+(\rep{16}_-,\rep{1})_\rep{-\frac32}$\\
& \!\!$4$ & $(\rep{8}_-,\rep{4})_\rep{-1}+(\rep{8}_-,\brep{4})_\rep{1}+(\rep{8}_+,\rep{6})_\rep{0}+(\rep{8}_+,\rep{1})_\rep{2}+(\rep{8}_+,\rep{1})_\rep{-2}$\\
& \!\!$5$ & $(\rep{4}_+,\rep{1})_\rep{\frac52}+(\rep{4}_-,\rep{1})_\rep{-\frac52}+(\rep{4}_+,\rep{5})_\rep{-\frac32}+(\rep{4}_-,\brep{5})_\rep{\frac32}+(\rep{4}_+,\brep{10})_\rep{\frac12}+(\rep{4}_-,\rep{10})_\rep{-\frac12}$\\
& \!\!$6$ & $(\rep{2}_+,\rep{1})_\rep{3}+(\rep{2}_+,\rep{1})_\rep{-3}+(\rep{2}_-,\rep{6})_\rep{-2}+(\rep{2}_-,\brep{6})_\rep{2}+(\rep{2}_+,\rep{15})_\rep{-1}+(\rep{2}_+,\brep{15})_\rep{2}$\\ 
 \hline
\end{tabular}
\end{center}
\caption{  \label{tab:branchingDN}
 Branching of the vector, adjoint, and spinor representations of D$_N$ into representations of $\text{D}_{n}\times \text{D}_{r}$ or $\text{D}_{n}\times \text{A}_{r-1}\times\text{U}(1)$ with $N=n+r$. 
 For the case where we have $\text{D}_2 = \text{A}_1\times \text{A}_1$, we denote $\rep{4} = (\rep{2}, \rep{2})$, $\rep{2}_+ = (\rep{2},\rep{1})$ and $\rep{2}_- = (\rep{1},\rep{2})$. The spinor representations are always Weyl spinors of SO($2N$). U(1) charges are given as subscripts on the representations. The associated bundle for these irreducible representations can be obtained as in the E$_8$ case.  
}
\end{table}

An important class of examples of this type of embeddings has been investigated by the authors of~\cite{Anderson:2011ns,Anderson:2012yf,Anderson:2013xka}. They consider bundles with structure group $\text{S(U(1)}{}^{r+1}\text{)} \subset \text{SU}(r+1)$ and write the bundle as a direct sum of line bundles~\eqref{eq:BundleDescription} with a trace constraint, 
\equ{
\label{eq:ExtendedBundleDescription}
\mathcal{V} = \bigoplus_{\tilde{a}=1}^{r+1}
 \mathcal{O}_X\big(k^{\tilde{a}}_1,\ldots, k^{\tilde{a}}_{h_{11}}\big)\,, 
\qquad 
\sum_{\tilde{a}=1}^{r+1} k^{\tilde{a}}_i = 0\,. 
}
The index $\tilde{a}$ labels the $r+1$ components of the vector representation of SU($r+1)$. 
This embedding into the ten-dimensional gauge group $\mathcal{G}$ is such that the $\tilde{a}^\text{th}$ line bundle is oriented along the $\tilde{a}^\text{th}$ Cartan direction\footnote{We label the these Cartan directions by $\tilde{a}$ rather than $I$, since they are associated to the bundle and not to the full Cartan subalgebra of the ten-dimensional gauge group, see Table~\ref{tb:IndexConventions}.} 
if the vectors $k_i$ are chosen sufficiently generic. (If this is not the case, enhancements of the unbroken gauge group are possible~\cite{Nibbelink:2015ixa}.)

We summarize the branching induced by this embedding in E$_8$ in Table~\ref{tab:extendedBranchingE8}. We list the branching of the adjoint $\rep{248}$ into irreducible representations of $G\times \widetilde{H}$ where $\widetilde{H}$ is the maximally enhanced SU($r$) structure group $H$ of the bundle. We also indicate for each irreducible representation the associated bundle. In Table~\ref{tab:branchingDN} we give some embeddings into SO($2N$) and list the branching of the relevant irreducible representations.

\subsubsection*{Identifying the irreducible representations after branching and computing the spectrum}

In the case where the bundle structure group embeds in $\text{A}_{r}$ it is possible to choose a U(1) basis such that each low energy irreducible representation that is paired with the fundamental representation of the bundle structure group is charged precisely under the $a^\text{th}$ Cartan generator. Note that the last bundle vector follows uniquely from the $r$ others due to the tracelessness condition~\eqref{eq:ExtendedBundleDescription}. On the level of the weights in the Dynkin basis this translates into the fact that the Dynkin labels of the $r$ states in the fundamental representation sum up to zero.

In order to determine the number of zero modes of an irreducible representation in the low energy theory one computes the index (or the dimensions of the appropriate sheaf cohomology groups) for the associated irreducible representation of the bundle structure group. For the extended branching of E$_8$ this means that, according to Table~\ref{tab:extendedBranchingE8}, in order to get the number of $\rep{10}$-plets of the low energy gauge group $G=\text{SU(5)}$, we have to compute the cohomology of the $\rep{5}$, i.e.\ of the fundamental representation of the bundle $\cV$. Likewise, for the number of $\brep{5}$-plets of $G$, we have to compute the cohomology of the $\rep{10}$, i.e.\ of the two-fold antisymmetrized representation of the bundle $\bigwedge^2 \cV$. Finally, the singlets of $G$ are paired with the adjoint of the bundle and hence one has to compute the multiplicity for $\cV \otimes \cV^*$. The computation of the dimensions of the relevant cohomology groups can be automated using the mathematica package developed in~\cite{Blumenhagen:2010pv,cohomCalg:Implementation}.

\subsubsection*{U(1)\texorpdfstring{$^{\boldsymbol{r}}$}{\^{}r} embeddings in $\boldsymbol{\mathcal{G}}$=SO($\boldsymbol{2N}$) }

For SO($2N$) the U(1)$^r$ structure group can be maximally enhanced to SO$(2r)$ or to SU$(r)\times U(1)$. We describe both embeddings into SO($2N$) in Table~\ref{tab:branchingDN}. We list for the former enhancement the branching of the relevant irreducible representations of SO($2N$), i.e.\ the vector, adjoint and spinor representation in full generality. For the latter embedding we give the branching of the spinor representations only for the spinors of SO($16$) that occur in the non-supersymmetric theory.

\subsubsection*{Redundancies and applicability}

In cases where the bundle structure group embeds into SU($r$) it is more natural to describe the bundle in terms of the Cartan subalgebra of this group. The description is also redundant (but much less so than the one of Section~\ref{sc:LeviEmbedding}) since it lists $r+1$ bundle charges for a bundle of rank $r$, i.e.\ the last bundle is uniquely determined from the tracelessness condition, which has to be imposed such that the bundle can embed into the (traceless) ten-dimensional gauge group. 

The embedding is defined implicitly by stating that it is such that the charges of the $r$ multiplets that transform in the fundamental representation of the structure group of the bundle carry precisely charge 1 under one of the Cartan generators and zero under all others. This choice is very natural since the bundle vectors in this basis correspond to the fundamental representation of $\cV$ and since it diagonalizes the charges of the individual bundle vectors in the Cartan subspace.

When computing the spectrum, this description has the advantage that the bundle vectors are chosen such that they describe (by definition) the fundamental representation of the structure group of $\cV$. Since for A$_r$ all irreducible representations can be constructed from the fundamental representation, all that remains to be done is to branch the ten-dimensional gauge group $\mathcal{G}$ into irreducible representations of the enhanced structure group $\widetilde{H}$ and the low energy gauge group $G$ and construct the other irreducible representations of $\widetilde{H}$ by taking symmetrized or anti-symmetrized products of the fundamental irreducible representation. In contrast, in the first description the bundle is written in terms of a linear combination of Cartan generators of $\mathcal{G}$. In particular, the charges are not such that the $r$ irreducible representations of $G$ that transform in the fundamental representation of $H$ carry charge 1 under precisely one Cartan generator. Consequently, the irreducible representations of $G$ in the spectrum computation cannot be found by simply taking appropriate powers of the line bundle vectors. Instead, one has to identify a weight vector of each irreducible representation of $G$ (usually one takes the highest weight, but this choice is irrelevant) and contract it with the bundle vectors in order to obtain the Chern classes of the divisors with respect to the Cartan directions that identify this state. Note that this is precisely what is done when calculating the spectrum via the Hirzebruch--Riemann--Roch index, which amounts to applying the multiplicity operator \eqref{MultiOp} to the root of $\mathcal{G}$ corresponding to the highest weight of the state $G$ in question.

However, in cases where the bundle structure group cannot be (fully) enhanced to a non-Abelian gauge group, the choice for the embedding in this description is not clear anymore. Of course it is possible to preform a combination of the descriptions of this and the previous sections: One identifies the part of the line bundle background that can be enhanced to non-Abelian factors for which one employs the description of this section. For the remaining Abelian directions one can use the Levi embedding as in Section~\ref{sc:LeviEmbedding}.

For bundles whose structure group enhances to SO($2r$), $\text{SU}(r)\times\text{U}(1)$ or even E$_r$, there is also no obviously preferable choice for the embedding. The lowest-dimensional representation of D$_r$ (the vector representation) has twice as many elements as the group has Cartan generators, such that an assignment in which each irreducible representation that pairs with the vector representation of the bundle has charge 1 under precisely one Cartan generator is not possible. For $\text{U}(1)^r\subset\text{SU}(r)\times\text{U}(1)$, one could choose the fundamental representation of SU($r$), but this will always be charged under the extra U($1$) in addition to the $r-1$ Cartan generators.

\section{Matching line bundle descriptions} 
\label{sc:MatchingBundles} 

In the previous sections we explained three different ways of describing the embedding of the bundle structure group $H$ into the ten-dimensional gauge group $\mathcal{G}$: 
\enums{ 
\item[{\bf 1.)}] Parameterize the line bundle background on a full Cartan subalgebra of the ten-dimensional gauge group (see Section~\ref{sc:AbelianFluxes}).  
\item[{\bf 2.)}] Use a Levi embedding of the line bundle background in the ten-dimensional gauge group (see Section~\ref{sc:LeviEmbedding}).
\item[{\bf 3.)}] Embed the line bundle background in a non-Abelian factor of a maximal subgroup of the ten-dimensional gauge group (see Section~\ref{sc:NonAbelianMaxEmbedding}).
}
In this section our aim is to relate the various descriptions to each other. From the discussions in the previous sections it is apparent that the different descriptions ultimately result from different choices of Cartan generators of the bundle. Consequently, they can be transformed into one another by a change of basis, as we briefly explain in the following.

\subsubsection*{Embedding 2.) $\boldsymbol{\rightarrow}$ Embedding 1.)}

Given a choice of simple roots associated to the nodes of the Dynkin diagram of $\mathcal{G}$ and the selection of the nodes that are deleted from the Dynkin diagram to define the Levi-embedding, we can express the gauge field strength as in~\eqref{eq:BundleFieldStrength}. We compare this to the general expansion~\eqref{LineBundleFlux} of the gauge field strength in terms of the Cartan generators $H_I$ of $\mathcal{G}$ and demand
\equ{ \label{MatchFieldStrengths} 
V_i^I\, H_I \stackrel{!}{=}  \ell_i^a\, h_a\,. 
}
From this we immediately obtain expressions of the line bundle vectors $V_i$ in terms of the bundle vectors $\ell_i$ in the Levi embedding, 
\equ{ \label{Matching:2->1} 
V_i^I = \ell_i^a \, (A^{-1})_{ab}\, \ga_b^I\,. 
}

\subsubsection*{Embedding 1.) $\boldsymbol{\rightarrow}$ Embedding 2.)}

Given a set of line bundle vectors $V_i$, we can use~\eqref{GaugeGroup} to determine the four-dimensional unbroken gauge group $G$. For this we can choose a set of simple roots and extend them to a set of simple roots of the ten-dimensional gauge group $\mathcal{G}$ by adding roots $\ga_a$. We then equate again the resulting field strengths in both descriptions as in~\eqref{MatchFieldStrengths}. By multiplying the resulting equation~\eqref{Matching:2->1} by $\ga_c^I$, we obtain 
\equ{ \label{eq:Emb1To2}
\ell_i^a = V_i^I \, \ga_a^I\,, 
}
using the definition of the Cartan matrix as in~\eqref{CartanMatrix}.

\subsubsection*{Embedding 1.) $\boldsymbol{\rightarrow}$ Embedding 3.)}

For simplicity we will only describe the often discussed case of line bundle backgrounds that can in principle be enhanced to SU($r+1$) given in~\eqref{eq:ExtendedBundleDescription}. We can choose a standard set of simple roots $\ga_a^{\tilde{a}}$, 
\equ{
\label{eq:ArCanonical}
\tga_1 = (1, -1, 0^{r-1})\,,
\qquad 
\tga_2 = (0,1,-1, 0^{r-2})\,,
\quad 
\ldots\,,
\quad 
\tga_{r} = (0^{r-1},1,-1)\,,
}
for the corresponding A$_r$ algebra whose roots are written as $(r+1)$-dimensional vectors labeled by $\tilde{a}=1,\ldots,r+1$ in the Cartan space. The extra degree of freedom in this embedding is removed via the tracelessness condition, which translates into 
\equ{
\sum_{\tilde{a}=1}^{r+1}\tga_a^{\tilde{a}}=0\,, 
}
for all $a$. 

Since not all line bundles can be enhanced to a semi-simple factor of a maximal subgroup of $\mathcal{G}$, this matching does not always work. (Of course, one can restrict to the sub-part of the line bundle background for which this is possible.) To avoid these complications, we will assume that the line bundle background can be enhanced to SU($r+1$) which is a semi-simple factor of a maximal subgroup of $\mathcal{G}$. 

We can proceed similarly to the  previous case, i.e.\ we start by finding a set of simple roots of the unbroken gauge group and complete it to a full set of roots of $\mathcal{G}$. Since the embedding in this case is via the extended Dynkin diagram, we construct from these roots a root system of the maximal semisimple subalgebra by adding the extended node and removing the appropriate other root. Now we define intermediate quantities $\kappa_i^a$, like the $\ell_i^a$ before, but with respect to the root set of the 10D algebra $\mathcal{G}$,
\begin{align}
\label{eq:defKappaCharges}
 \kappa_i^a=V_i^I\,  \alpha_a^I,
\end{align}
for all broken roots $\alpha_a$. Depending on the embedding, this might or might not involve the extended root. The $k_i^a$ are defined with respect to the standard choice of the SU($r+1$) roots in~\eqref{eq:ArCanonical} such that they are charged under precisely one Cartan generator $\tilde{h}^{\tilde{a}}$. As in \eqref{LeviCartanBasis}, this is enforced via the inverse Cartan matrix: Defining $\widetilde{A}^{-1}$ as the inverse Cartan matrix of the Lie algebra A$_r$, we find
\equ{
\label{eq:Emb1To3}
 k_i^{\tilde{a}} = \kappa_i^a\, (\widetilde{A}^{-1})_{ab} \,\tga_b^{\tilde{a}}\,. 
}

\subsubsection*{Embedding 3.) $\boldsymbol{\rightarrow}$ Embedding 1.)}

Embedding 1.) can be obtained by inverting the steps outlined above. We first note that \eqref{eq:Emb1To3} can be inverted by multiplying with $\tga_c^{\tilde{a}}$,
\begin{align}
 \kappa_i^a=k_i^{\tilde{a}}\, \tga_a^{\tilde{a}}\,.  
\end{align}
Subsequently, we invert relation \eqref{eq:defKappaCharges} by multiplying with $(A^{-1})_{ab}\alpha_b^I$,
\begin{align}
 \label{eq:Emb3To1}
 V_i^I=\kappa_i^a \,(A^{-1})_{ab}\,\alpha_b^I\,.  
\end{align}

In~\cite{Nibbelink:2015ixa,Blaszczyk:2015zta} a way to translate the S(U(1)$^{5}$) line bundles into the language of line bundle vectors was presented. (For completeness we have added that method in Appendix~\ref{sc:IdUbundles}.)

\subsubsection*{Embedding 2.) $\boldsymbol{\rightarrow}$ Embedding 3.) and vice versa}

Obviously, this matching can be performed by first matching embedding 2.) to embedding 1.) and then matching embedding 1.) to embedding 3.), and vice versa. Since all maps are linear, given by appropriate matrix multiplications, these steps can also be combined.

\subsubsection*{Comparison of the three embeddings}

A big advantage of the description of Section~\ref{sc:LeviEmbedding} over the one described in Section~\ref{sc:NonAbelianMaxEmbedding} is that it generalizes in a straightforward manner to cases where the bundle does not embed into S(U(1)$^{r}$). As the method describes explicitly the embedding of the bundle into the Cartan space of $\mathcal{G}$, it is irrelevant whether or not the bundle structure group can be enhanced to a non-Abelian group. In contrast, the latter description crucially relies on this fact, since it describes the embedding only implicitly by following the convention that the Cartan generators of the bundle are oriented in the Cartan space of the ten-dimensional gauge group such that each of the $r$ states that transform in the fundamental irreducible representation of the bundle is charged under exactly one U(1). In cases where the bundle does not embed into a larger structure group, the embedding thus needs to be fixed by some other means. Compared to the line bundle vectors introduced in Section~\ref{sc:AbelianFluxes} the description of Section~\ref{sc:LeviEmbedding} is much less redundant since the vectors $\ell_i$ have only rank($H$) components, whereas the line bundle vectors $V_i$ always have 16 components. The description of the bundle in Section~\ref{sc:NonAbelianMaxEmbedding} does not rely on an explicit choice of the embedding of the structure group into the root system of $\mathcal{G}$ which makes this description much less redundant.

\section{Examples}
\label{sc:examples} 

In this section we present three examples with different characteristics in order to illustrate the various embeddings and how to transform them into one another.

\subsection[Example: Embedding S(U(1)\texorpdfstring{$^{5}$}{\^{}5}) line bundles in an E\texorpdfstring{$_{8}$}{\_{}8} factor]{Example: Embedding S(U(1)\texorpdfstring{$^{\boldsymbol{5}}$}{\^{}5}) line bundles in an E\texorpdfstring{$_{\boldsymbol{8}}$}{\_{}8} factor}

Let us use start with an example of a bundle $\mathcal{V}$ with structure group $H=\text{S}(\text{U}(1)^{5})\subset\text{SU}(5)$. These occur frequently due to their phenomenological relevance. For concreteness we will work with $h_{11}=4$ in this example. Using embedding 1.), we can describe $\mathcal{V}$ via the bundle vectors
\begin{align}
 V_i^{\prime I}=(a^I_i,a^I_i,a^I_i,a^I_i,a^I_i,b^I_i,c^I_i,d^I_i)\,,
 \quad 
 V_i^{\prime\prime} = 0\,,
\end{align}
where $i=1,\ldots,4$ labels the four divisors and $I=1,\ldots,8$ labels the 8 Cartan generators of the observable E$_8$. This U(1)$^4$ bundle has already been investigated in \cite{Blaszczyk:2015zta} (see Appendix~\ref{sc:IdUbundles} for a review of the method used there). For sufficiently generic $a_i,b_i,c_i$, this will leave the four simple roots
\begin{align}
\label{eq:Ex1SRTUnbroken}
\arry{l}{
 \alpha_8=(1,-1,0,0,0,0,0,0)\,,\\ \alpha_5=(0,1,-1,0,0,0,0,0)\,,
 }
 \qquad 
 \arry{l}{
 \alpha_6=(0,0,1,-1,0,0,0,0)\,,\\ \alpha_7=(0,0,0,1,-1,0,0,0)\,,
 }
\end{align}
unbroken, where we have already assigned the standard numbering of simple roots in E$_8$. For computing the spectrum, one could now take the inner product of these $\alpha_a$ with all 248 roots $\lambda$ of E$_8$ to identify the irreducible representation of which $\lambda$ is the (highest) weight and apply the multiplicity operator to these states.

\subsubsection*{Transformation to embedding 2.)}

In order to transform the basis of the $V_i^I$ into the basis of the $\ell_i^a$, we first need to find those simple roots $\alpha_a$ of E$_8$ that are broken by the $V_i^I$. In order to find these $\alpha_a$ we successively add E$_8$ roots that have the correct inner product relations such that the Cartan matrix of E$_8$ is reproduced. 
A possible choice for the four additional simple roots of E$_8$ are: 
\begin{align}
\label{eq:Ex1SRTBroken}
\arry{l}{
\alpha_1 =(0, 0, 0, 0, 0, 1, -1, 0)\,, \\ \alpha_2=(0, 0, 0, 0, 0, 0, 1, -1)\,,
 }
 \qquad 
 \arry{l}{
 \alpha_3 =\left(\sfrac12, \sfrac12, \sfrac12, \sfrac12, \sfrac12, -\sfrac12, -\sfrac12, \sfrac12\right)\,,\\ 
 \alpha_4=(-1, -1, 0, 0, 0, 0, 0, 0)\,.
 } 
\end{align}
Using \eqref{eq:Emb1To2}, we find
\begin{align}
\ell_i^a = \big(b_i - c_i,  c_i - d_i, \sfrac12( 5 a_i - b_i - c_i + d_i), -2 a_i\big)\,.
\end{align}
These are four vectors with four components, or a 4$\times$4 matrix. The index $a$ labels the four line bundles associated with $\alpha_a$. The index $i$ labels the divisors $D_i$, $i=1,\ldots,h_{11}$, which is also four in this example. Reading the matrix as vectors labeled by $a$ whose components are labeled by $i$, each vector corresponds to a line bundle and its components are the first Chern classes of the four divisors. Reading it the other way, the four vectors correspond to the four divisors and the four entries describe the contribution from the four Line bundles to the first Chern class of the divisor.

\subsubsection*{Transformation to embedding 3.)}

From the combined root system \eqref{eq:Ex1SRTUnbroken} and \eqref{eq:Ex1SRTBroken} of unbroken and broken roots, respectively, we construct the extended root system and exchange the root $\alpha_4$ for the extended root $\alpha_0$ using relation~\eqref{eq:ExtendedRootRelation}. We then compute the $\kappa_i^a$ according to~\eqref{eq:defKappaCharges} and find
\begin{align}
 \kappa_i^a=\big(c_i+d_i,b_i-c_i,c_i-d_i,\sfrac12(5a_i-b_i-c_i+d_i)\big)\,.
\end{align}
After the second step \eqref{eq:Emb1To3} we find for the $k_i$
\begin{align}
\arry{l}{ 
 k^1_i=\frac{a_i}{2}+\frac{b_i}{2}+\frac{c_i}{2}+\frac{d_i}{2}\,, \\ k^2_i=\frac{a_i}{2}+\frac{b_i}{2}-\frac{c_i}{2}-\frac{d_i}{2}\,,
 }
\qquad 
\arry{l}{ 
 k^3_i=\frac{a_i}{2}-\frac{b_i}{2}+\frac{c_i}{2}-\frac{d_i}{2}\,, \\  k^4_i=\frac{a_i}{2}-\frac{b_i}{2}-\frac{c_i}{2}+\frac{d_i}{2}\,,
}
 \qquad  
 k^5_i =-2 a_i\,.
\end{align}
Note that all $k_i^{\tilde{a}}$ automatically sum to zero (summing over $\tilde{a}$ for fixed $i$) as dictated by the tracelessness condition on the U(1) charges. As before, the matrix $k_i^{\tilde{a}}$ can be read column-wise or row-wise. Note that we have now added an extra (linearly dependent) root, such that $\tilde{a}$ runs from 1 to 5 now. The results match with those of Appendix \ref{sc:IdUbundles} up to a permutation of the individual line bundles and an overall minus sign, which is not fixed since it is a matter of convention whether one considers $\cV$ or $\cV^*$.

From these U(1) bundles we can now easily construct the bundles associated with any irreducible representation of the low energy gauge group $G$ in order to compute the number of zero modes. In particular, the $\brep{5}$-plets of $G$ pair up with the $\rep{10}$-plet of $H$ (which is the two-fold antisymmetrized fundamental representation), such that the ten distinct $\rep{5}$-plets are given by the 10 bundle vectors
\begin{align}
 \mathcal{L}^{\tilde{a},\tilde{b}}=\mathcal{O}_X(k^{\tilde{a}}_i + k^{\tilde{b}}_i)\,,
\end{align}
with $\tilde{a}<\tilde{b}$, and likewise for the other representations.

For computing traces we note that in the vector representation we have 
\begin{align}
 \tr\,H^{\tilde{a}}H^{\tilde{b}}=\delta^{\tilde{a}\tilde{b}}
\end{align}
and consequently the Chern characters are obtained with the standard scalar product between the $k^{\tilde{a}}_i$.

\subsection[Example: Embedding U(1)\texorpdfstring{$^{2}$}{\^{}2} line bundles in an E\texorpdfstring{$_{8}$}{\_{}8} factor]{Example: Embedding U(1)\texorpdfstring{$^{\boldsymbol{2}}$}{\^{}2} line bundles in an E\texorpdfstring{$_{\boldsymbol{8}}$}{\_{}8} factor}

In this example we consider a U(1)$^2$ line bundle that embeds into the first E$_8$ factor such that the unbroken gauge group is SU(5)$\times$SU(3), i.e.\ the breaking proceeds by deleting the first and fourth node of the Dynkin diagram. For the sake of this example we explicitly discuss the following line bundle on a CY with $h_{11}=3$:
\begin{align}
 V^\prime_1=\left(\sfrac{1}{2}, \sfrac{1}{2}, -\sfrac{1}{2}, -\sfrac{1}{2}, \sfrac{1}{2}, \sfrac{1}{2}, \sfrac{1}{2}, \sfrac{1}{2}\right)\,, 
 \qquad 
 \arry{l}{
 V^\prime_2=\left(-\sfrac{3}{2}, -\sfrac{3}{2}, \sfrac{1}{2}, \sfrac{3}{2}, -\sfrac{5}{2}, ~~\sfrac{1}{2}, -\sfrac{1}{2}, -\sfrac{5}{2}\right)\,,
\\
 V^\prime_3=\left(-\sfrac{1}{2}, -\sfrac{1}{2}, \sfrac{3}{2},\sfrac{1}{2}, ~~\sfrac{1}{2}, -\sfrac{5}{2}, -\sfrac{3}{2}, ~~\sfrac{1}{2}\right)\,, 
 }
\end{align}
and $V_1^{\prime\prime} = V_2^{\prime\prime} = V_3^{\prime\prime} = 0$. This is a U(1)$^2$ bundle rather than a U(1)$^3$ bundle since it satisfies $4V_1+V_2+V_3=0$, i.e.\ the three line bundles on the three divisors are not linearly independent. Using the techniques outlined in Appendix~\ref{sc:Branchings}, we find the projection matrix
\begin{equation}
\label{eq:SU5xSU3BranchingMatrix}
 P_{E_8\subset \text{A}_4\times \text{A}_3\times \text{U}(1)^2}=
 \left(
 \scalebox{.6}{$
 \begin{matrix}
 2 & 3 & 4 & 5 & 6 & 4 & 2 & 3 \\ \hline 
 0 & 1 & 0 & 0 & 0 & 0 & 0 & 0 \\
 0 & 0 & 1 & 0 & 0 & 0 & 0 & 0 \\ \hline 
 5 & 10 & 15 & 20 & 24 & 16 & 8 & 12 \\ \hline 
 0 & 0 & 0 & 0 & 0 & 0 & 0 & 1 \\
 0 & 0 & 0 & 0 & 1 & 0 & 0 & 0 \\
 0 & 0 & 0 & 0 & 0 & 1 & 0 & 0 \\
 0 & 0 & 0 & 0 & 0 & 0 & 1 & 0 \\ 
 \end{matrix}$}
 \right)\,,
 \end{equation}
where the rows that correspond to the two U(1) charges have been separated by lines. This yields the following branching of the adjoint of E$_8$ into SU(5)$\times$SU(3)$\times$U(1)$^2$ (U(1) charges are written as subscripts):
\begin{align}
\label{eq:SU5xSU3Split}
\begin{split}
\rep{248}\rightarrow &\big[(\rep{10},\rep{3})_{0,1}+ (\rep{10},\rep{1})_{-1,-4} +(\rep{10},\rep{1})_{1,1}+(\rep{1},\rep{3})_{-1,0}+(\rep{1},\brep{3})_{-1,-5}+(\rep{5},\rep{3})_{0,-2}+(\rep{5},\brep{3})_{1,3}\\
&+(\brep{5},\rep{3})_{1,2}+(\rep{5},\rep{1})_{0,3} +(\rep{1},\rep{1})_{2,5} +\text{h.c.}\big]+ (\rep{24},\rep{1})_{0,0}+(\rep{1},\rep{8})_{0,0}+(\rep{1},\rep{1})_{0,0}+(\rep{1},\rep{1})_{0,0}
\end{split}
\end{align}
The last four terms are the adjoints of the four low energy gauge groups and the rest constitute the matter content. As a simple crosscheck one finds that the spectrum after branching is non-chiral and that the dimensions of the representations add up to $248$ as it should be when branching the adjoint of E$_8$.

It is also instructive to look at the branching induced by each individual line bundle vector $V_i'$, for which we find
\begin{align}
\label{eq:SU5xSU3InducedBranching}
\begin{split}
 V_1^\prime: \text{E}_8&\rightarrow \text{E}_7\times \text{U(1)}\,,\\
 V_2^\prime: \text{E}_8&\rightarrow \text{SU(5)}\times\text{SU(4)}\times \text{U(1)}\,,\\
 V_3^\prime: \text{E}_8&\rightarrow \text{SO(10)}\times \text{SU(3)}\times\text{U(1)}\times\text{U(1)}\,. 
\end{split}
\end{align}
In particular, $V^\prime_1$ breaks the first root and $V^\prime_2$ breaks the fourth root. The vector $V^\prime_3$ breaks a linear combination of the two such that in the end one obtains only two U(1) factors.

In order to compute the matter spectrum from here one now has to contract the E$_8$ roots that correspond to the (highest weight of the) matter irreducible representation in question with the line bundle vectors $V^\prime_i$. Thus, in order to find the number of $(\rep{10},\rep{3})_{0,1}$ states one applies the corresponding root $\lambda$ to the multiplicity operator \eqref{MultiOp} (or equivalently uses  the Hirzebruch--Riemann--Roch index on the inner product of the bundle vectors with $\lambda$). 

In order to proceed, we first find a basis for the simple roots of the unbroken gauge groups. We choose
\begin{align}
\arry{l}{
\alpha_2=(0,0,1,0,0,0,1,0)\,,
\\ 
\alpha_5=(0,0,0,0,1,0,0,-1),
\\ 
\alpha_7=(0,1,0,1,0,0,0,0)\,,
}
\qquad 
\arry{l}{
\alpha_3=(\sfrac12,\sfrac12,-\sfrac12,-\sfrac12,-\sfrac12,-\sfrac12,-\sfrac12,-\sfrac12)\,,
\\
\alpha_6=(\sfrac12,-\sfrac12,\sfrac12,-\sfrac12,-\sfrac12,\sfrac12,-\sfrac12,\sfrac12)\,,
\\
\alpha_8=(\sfrac12,-\sfrac12,-\sfrac12,\sfrac12,-\sfrac12,-\sfrac12,\sfrac12,\sfrac12)\,.  
}
\end{align}
The numbering has again been chosen according to the standard numbering convention within E$_8$. The roots $(\alpha_2,\alpha_3)$ are the simple roots of SU(3) and the roots $(\alpha_8,\alpha_5,\alpha_6,\alpha_7)$ are the simple roots of SU(5). This set of roots can be completed to a set of simple roots of E$_8$ by adding the roots
\begin{align}
\label{eq:Example2BrokenRoots}
\alpha_1=(0,0,-1,0,0,1,0,0)\,,
\qquad
\alpha_4=(-1,0,0,0,0,0,0,1)\,.
\end{align}

Using \eqref{eq:SU5xSU3BranchingMatrix}, one finds the two U(1) generators $T_A$: $T_1^I=V_1^{\prime I}$ and $T_2^I=-V_2^{\prime I}$. (We use labels $A,B \in \{1,2\}$ to distinguish them from the labels $a,b$ since we have chosen a specific basis here). This corresponds to the fact that the $V_i'$ parameterize the direction of broken Cartan generators and can consequently be used as U(1) generators. It is instructive to discuss an embedding of the $V^I_i$ into $\E8$ using these two generators, even though this actually corresponds to neither embedding 2 nor 3 discussed above. We will also carry out the construction of embedding 2 explicitly and point out its close relation to the embedding via the $T$'s. Note that embedding 3 cannot be constructed for this bundle since it is not a simple S(U(1)$^3$) bundle due to the more complicated sum relation this bundle satisfies. In the description of the bundle with respect to the two $\U1$ generators $T_A$ we call the two three-component line bundles $\mathcal{L}^A=\mathcal{O}_X(p^A_i)$, where the three components correspond to the Chern classes of the three divisors $D_i$, $i=1,\dots, h_{11}=3$.

In this particular example the E$_8$ root corresponding to the highest weight of the $(\rep{1},\rep{3})_{-1,0}$ is
\begin{align}
 \lambda_{(1,3)}=(-\sfrac12,-\sfrac12,\sfrac12,\sfrac12,\sfrac12,-\sfrac12,-\sfrac12,\sfrac12)\,,
\end{align}
hence its charges read $\lambda_{(1,3)}\cdot V^\prime = (-1,0,4)$ with respect to the divisors $D_1,D_2,D_3$ and consequently the corresponding line bundle $\mathcal{L}^1$ reads
\begin{align}
 \mathcal{L}^1 =\mathcal{O}_X(-1,0,4)\,.
\end{align}
Likewise, we find for the highest weight of the $(\rep{10},\rep{3})_{0,1}$,
\begin{align}
 \lambda_{(10,3)}=(-1,0,0,0,0,0,0,1)\,,
\end{align}
the charges $\lambda_{(10,3)}\cdot V^\prime = (0,-1,1)$ and thus
\begin{align}
 \mathcal{L}^2 = \mathcal{O}_X(0,-1,1)\,.
\end{align}
Since $T_1$ is oriented along $V_1'$ (which parameterizes the line bundle at the first divisor) and $T_2$ along $V_2'$ (which parameterizes the line bundle at the second divisor), the first and second entries of $\mathcal{L}^1$ parameterize loosely speaking how much of the first Chern classes of the flux on the divisor $D_1$ come from $V_1'$ and $V_2'$, respectively. Likewise, $\mathcal{L}^2$ parameterizes the contributions of $V_1'$ and $V_2'$ to the first Chern classes of the flux on the divisor $D_2$. In this choice, the flux on $D_1$ is entirely due to $V_1'$ and the flux on $D_2$ is entirely due to $V_2'$. $V_3'$ contributes to both since $4V_1'+V_2'+V_3'=0$.

Using $\mathcal{L}^1$ and $\mathcal{L}^2$, one can construct any other irreducible representation. For example we know that the singlets $(\rep{1},\rep{1})_{2,5}$ pair up with the the two U(1) bundles such that their charges are $(2,5)$ and consequently
\begin{align}
 (\rep{1},\rep{1})_{2,5} \quad\leftrightarrow\quad (\mathcal{L}^{1*})^{2} \times  (\mathcal{L}^2)^{5}=\mathcal{O}_X(2,-5,-3)\,,
\end{align}
where $\mathcal{L}^*$ denotes to dual bundle of $\mathcal{L}$. Of course we find the same result from contracting the corresponding root
\begin{align}
 \lambda_{(1,1)}=\left(\sfrac{1}{2},\sfrac{1}{2},-\sfrac{1}{2},-\sfrac{1}{2},\sfrac{1}{2},\sfrac{1}{2},\sfrac{1}{2},\sfrac{1}{2}\right)
\end{align}
with $V_1', V_2' ,V_3'$, respectively.

While this basis is convenient for computing various matter representations via index theorems or sheaf cohomology one has to be careful when computing the second Chern character of the bundle. In contrast to the case where the bundle structure group embeds into a non-Abelian SU($N$) group, where the Cartan directions are chosen orthonormal, the U(1) generators $T_1$ and $T_2$ we have chosen here are neither normalized nor orthogonal. This means that the second Chern character of the bundle is not simply given by the square of the $\mathcal{L}_i^A$'s contracted with the divisors. Rather, using $\tr{H^I H^J}=\delta^{IJ}$ one finds that 
\begin{align}
 V_1'^2=2\,,\quad  V_1' V_2'=-5\,,\quad V_2'^2=20\,.
\end{align}
Using the relation $4V_1'+V_2'+V_3'=0$ the other scalar products follow. Thus, the second Chern character of the bundle is given by
\begin{align}
 \label{eq:ChernCharacterChargeDiagBasis}
 \text{ch}_2(\mathcal{V})= -\frac 12\, p_i^A G_{AB} p_j^B D_i D_j\,, \qquad G_{AB}=\begin{pmatrix}2&-5\\-5&20\end{pmatrix}\,,
\end{align}
The resulting expression matches exactly the second Chern character as computed directly from the $V_i'$ via \eqref{eq:secondChernCharacterBundle}.
Alternatively, the same result can be obtained from explicitly summing up the traces for the branching \eqref{eq:SU5xSU3Split}, i.e.\ by computing
\begin{align}
 \sum_{i,j=1}^{2} q_i \,q_j~ \text{dim}(\mathbf{R}_{(q_i,q_j)})\,,
\end{align}
where $\text{dim}(\mathbf{R}_{(q_i,q_j)})$ is the dimension of the irreducible representations appearing in \eqref{eq:SU5xSU3Split}. Note that $\tr\cF^2$ is related to the U(1) charges since the U(1) generators correspond to the bundle vectors $V'$ as explained below \eqref{eq:Example2BrokenRoots}.

In order to compute the bundle in the Levi embedding basis $\ell_i^a$ we use \eqref{eq:Emb1To2}, which yields
\begin{align}
 \ell^1=(1,0,-4)\,,
 \quad 
 \ell^4=(0,-1,1)\,.
\end{align}
The $h_a^I$ are found via \eqref{LeviCartanBasis},
\begin{align}
 h_1^I=V_1'^I\,,
 \quad 
 h_2^I=V_2'^I\,,
\end{align}
i.e.\ they are, up to a sign, equal to the U(1) generators from above. Consequently, the $\ell^a$ are also closely related to the two bundles $\mathcal{L}^1$ and $\mathcal{L}^2$ we have also encountered above, since they were chosen such that they are also only charged under one of the U(1)s. The only difference\footnote{Note that it is a mere matter of convention whether one considers $\mathcal{V}$ or $\mathcal{V}^*$.} is that $\ell^{1}=-\mathcal{L}^1$. As for the Chern characters, we recognize the components of the matrix $G_{AB}$ above as the entries $(A^{-1})_{ab}$ of the inverse Cartan matrix of E$_8$ for 
$A,B=1,2$ corresponding to $a,b=1,4$, respectively: $G_{AB} = (-)^{a+b}\, (A^{-1})_{ab}$. The extra minus signs in $G_{12}$ and $G_{21}$ in \eqref{eq:ChernCharacterChargeDiagBasis} is due to the relative minus signs between $\ell^1$ and $\mathcal{L}^1$.

\subsection[Example: Embedding U(1)\texorpdfstring{$^{3}$}{\^{}3} line bundles in SO(16)]{Example: Embedding U(1)\texorpdfstring{$^{\boldsymbol{3}}$}{\^{}3} line bundles in SO(16)}

In this final example we want to present the embedding of a sum of line bundles in an SO group. As an example we investigate the embedding for one SO(16) factor of the bundle discussed with $h_{11}=4$ in \cite{Blaszczyk:2015zta}. This bundle belongs to the non-supersymmetric theory which, in contrast to the supersymmetric $\text{E}_8\times\text{E}_8$, carries bi-fundamental matter that is charged under both SO(16) factors simultaneously. Such a bundle, if non-trivial in both sectors cannot simply be split and we do this here only for ease of exposition. Since this example is very similar to our first one we will be rather brief. The bundle vectors are
\begin{align}
\label{eq:SO16SU5U14Branching}
\arry{l}{
V^\prime_1 = (-1, ~~1, ~~2, -1, -1, -1, ~~2, ~~1)\,, \\ V^\prime_2 = (~~0, -1, -1, ~~0, ~~0, ~~0, ~~0, ~~0)\,,
}
\quad 
\arry{l}{
V^\prime_3 = (~~0, ~~1, ~~1, ~~0, ~~0, ~~0, -2, ~~0)\,,\\ V^\prime_4 = (~~1, ~~0, -1, ~~1, ~~1, ~~1, ~~0, -1)\,, 
}
\end{align}
and $V^{\prime\prime}_1 = \ldots = V^{\prime\prime}_4 = 0$. 
The bundle satisfies $V_1+2V_2+V_3+V_4=0$ and is hence a $\text{U}(1)^{3}$ bundle. We will focus on the branching of the adjoint. The spectrum in terms of $\text{SU}(5) \times U(1)^{4}$ is found via the projection matrix
\begin{align}
P_{\text{D}_8\subset \text{A}_4\times \text{U}(1)^4}=\left(
\scalebox{.6}{$
\begin{array}{cccccccc}
 1 & 0 & 0 & 0 & 0 & 0 & 0 & 0 \\
 0 & 1 & 0 & 0 & 0 & 0 & 0 & 0 \\
 0 & 0 & 1 & 0 & 0 & 0 & 0 & 0 \\
 0 & 0 & 0 & 1 & 0 & 0 & 0 & 0 \\ \hline 
 1 & 2 & 3 & 4 & 5 & 5 & \frac{5}{2} & \frac{5}{2} \\
 1 & 2 & 3 & 4 & 5 & 6 & 3 & 3 \\
 \frac{1}{2} & 1 & \frac{3}{2} & 2 & \frac{5}{2} & 3 & 2 & \frac{3}{2} \\
 \frac{1}{2} & 1 & \frac{3}{2} & 2 & \frac{5}{2} & 3 & \frac{3}{2} & 2 \\
\end{array}$}
\right)\,,
\end{align}
which yields the following branching of the adjoint $\rep{120}$ of SO(16):
\begin{align}
 \begin{split}
  120\rightarrow\big[&\rep{10}_{2211}+\rep{5}_{1000}+\rep{5}_{1100}+\rep{5}_{1110}+\rep{5}_{1101}+\rep{5}_{1111}+\rep{5}_{1211}\\
  &+\rep{1}_{0001}+\rep{1}_{0010}+\rep{1}_{0100}+\rep{1}_{0110}+\rep{1}_{0101}+\rep{1}_{0111}+\text{h.c.}\big]+\rep{24}_{0000}+ 4~\rep{1}_{0000}\,.
 \end{split}
\end{align}
Similarly, this can be obtained from enhancing U(1)$^4$ to SU(4)$\times$U(1) and embedding this into SO(16) using that $\text{SU(4)}\simeq\text{SO}(6)$ as a Lie algebra. Using Table~\ref{tab:branchingDN} one finds for the adjoint
\begin{align}
\label{eq:SO16SU4Branching}
 120\rightarrow \big[(\rep{10},\rep{1})_{2}+(\rep{5},\rep{6})_1+\text{h.c.}\big]+(\rep{24},\rep{1})_0+(\rep{1},\rep{15})_0+(\rep{1},\rep{1})_0\,.
\end{align}
The corresponding projection matrix is given by replacing the fifth root by the fifth row of the inverse Cartan matrix of SO(16). The spectrum \eqref{eq:SO16SU4Branching} makes the origin of the $\rep{10}$-plet, the six $\rep{5}$-plets (plus their charge conjugates) as well as the total number of 12 charged singlets in \eqref{eq:SO16SU5U14Branching} evident. We choose the following simple roots
\begin{align}
 \arry{l}{
  \alpha_1=(1, 0, 0, -1, 0, 0, 0, 0)\,, \\ \alpha_2=(0, 0, 0, 1, -1, 0, 0, 0)\,,
  }
  \quad 
  \arry{l}{
  \alpha_3= (0, 0, 0, 0, 1, -1, 0, 0)\,,\\ \alpha_4=(0, 0, 0, 0, 0, 1, 0, 1)\,, 
  }
\end{align}
as the simple roots of the unbroken SU(5) and complete them to the simple roots of SO(16) with the four roots $\ga_a$: 
\begin{align}
 \arry{l}{
  \alpha_5=(0, -1, 0, 0, 0, 0, 0, -1)\,, \\ \alpha_6=(0, 1, -1, 0, 0, 0, 0, 0)\,,
  }
  \quad 
  \arry{l}{
  \alpha_7=(0, 0, 1, 0, 0, 0, 1, 0)\,, \\ \alpha_8=(0, 0, 1, 0, 0, 0, -1, 0)\,.
 }
\end{align}
Since the bundle is only a U(1)$^3$ bundle, one linear combination of U(1)s will stay massless. With this choice one finds for the $\ell_i^a$
\begin{align}
 \ell^5=(-2, -1, 4, 0)\,,~\ell^6=(1, 0, -1, -1)\,,~\ell^7=(-1, 0, -1, 3)\,,~\ell^8=(1, 1, -1, -1)\,.
\end{align}
Similarly to the last example, the entries satisfy $\ell^5_i+2\ell^6_i+\ell^a_i+\ell^8_i=0$ for all $i$ due to the equivalent relation among the $V_i'$. When computing the second Chern character one has to take the inner product with respect to the matrix $G_{ab}=(A^{-1})_{ab}$, $a,b=5,6,7,8$.

\section*{Acknowledgments}
We thank Ralph Blumenhagen for providing motivation to work on this project and Orestis Loukas for providing technical details for the examples that are discussed in this paper. 
SGN would like to thank the Center for Theoretical Physics at the Sichuan University for their kind hospitality. 
The work of F.R.\ is supported by the German Science Foundation (DFG) within the Collaborative Research Center (SFB) 676 ``Particles, Strings and the Early Universe''.

\bigskip

\begin{appendices}

\def\theequation{\thesection.\arabic{equation}} 
\setcounter{equation}{0}
\section{Some elements of group theory}
\label{sc:GroupTh}

In the following Appendix we review some of the group-theoretical methods that can be employed to describe the branching of gauge groups and their representations. Details can be found e.g.\ in~\cite{Slansky:1981yr,Fuchs:1997jv}. We focus on the cases of interest to heterotic string theory, namely $\mathcal{G}= \text{E}_8$ and SO(2$N$).

\subsection{Simple roots and Cartan matrix}

\begin{figure}[t]
     \centering
     \subfloat[][Extended Dynkin diagram of E$_8$]{\includegraphics[width=.55\textwidth]{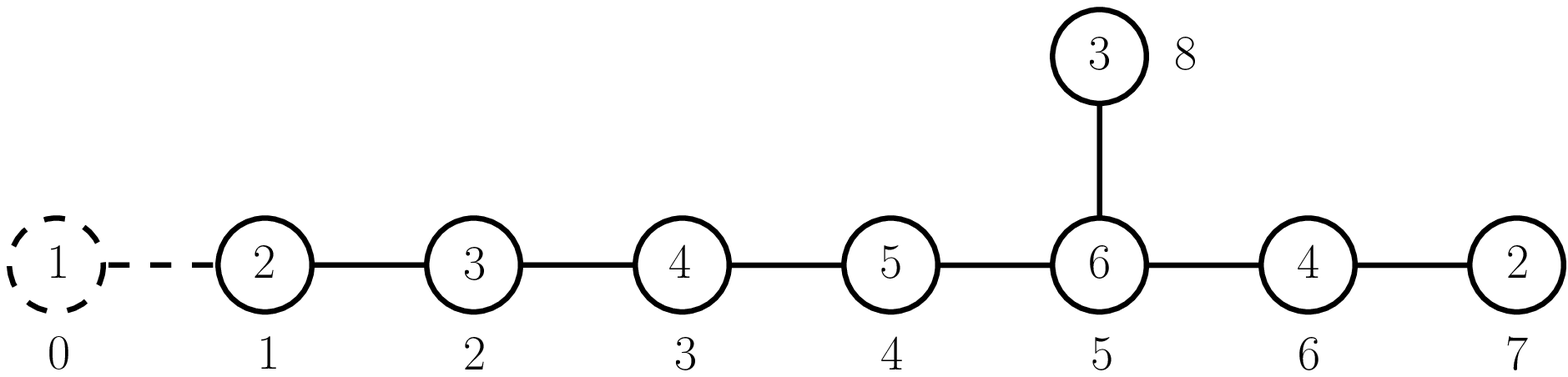}\label{fig:ExtendedE8DD}}
     \qquad
     \subfloat[][Extended Dynkin diagram of SO($2N$)]{\includegraphics[width=.35\textwidth]{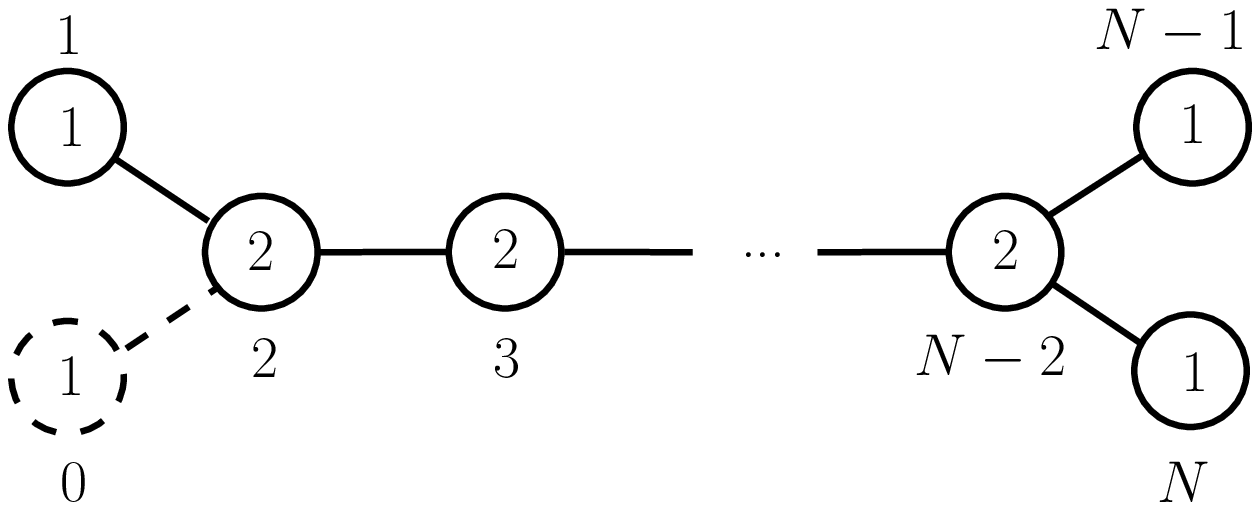}\label{fig:ExtendedSO2NDD}}
      \caption{      \label{fig:ADEDynkin}
      The numbers outside the nodes indicate our numbering convention for simple roots. The number inside the nodes are the Dynkin multiplicities. The dashed node is the extended root of the extended Dynkin diagram.}
\end{figure}

The roots of the Lie algebras E$_8$ and SO(2$N$) are summarized in Table \ref{tb:10DHetGroups}. For these roots one can choose a notion of positivity, dividing the non-zero roots into positive and negative sets. From the positive roots one can define rank $r$ simple roots $\alpha_a$, from which all other positive (negative) roots can be constructed by adding simple roots with non-negative (non-positive) coefficients. While the choice of positivity and consequently the set of simple roots is arbitrary, the Cartan matrix
\begin{align} \label{CartanMatrix} 
 A_{ab}=\frac{2\,\alpha_a\cdot\alpha_b}{\alpha_a\cdot\alpha_a} = \ga_a\cdot\ga_b~,
\end{align}
where the dot denotes the standard Euclidean scalar product, is independent of this choice. The second equality is only true for simply-laced algebras; but those are the only ones that we are concerned with here. Of course the labelling of the roots is arbitrary and we summarize our conventions for the groups E$_8$ and SO($2N$) in Figure~\ref{fig:ADEDynkin}.

As in the case of the Cartan matrix, one can work with Dynkin labels rather than the root vectors in order to not depend on an explicit choice of simple roots, positivity, and so on. The Dynkin labels $a_a$ associated with a root $\lambda$ are determined via its scalar product with the simple roots,
\begin{align}
 a_a(\lambda) = \lambda\cdot\alpha_a\,.
\end{align}
The highest root is the root whose Dynkin labels are all non-negative. Minus the highest root is called $\alpha_0$. As all roots, it is a linear combination of simple roots,
\begin{align}
\label{eq:ExtendedRootRelation}
 c_0\,  \alpha_0 + \sum c_a\, \alpha_a=0\,. 
\end{align}
The $c_a$ are referred to as Dynkin multiplicities and are normalized such that $c_0=1$. This root is very important in the classification of subalgebras and is added as an extended root to the Dynkin diagram or the Cartan matrix.

\subsection{Irreducible representations and highest weights}

The Dynkin labels can be defined for any irreducible representation by computing the inner product of the simple roots with the corresponding weights $w$ instead of the roots $\lambda$. This is called the Dynkin basis. Each irreducible representation is uniquely identified by its highest weight $W$. The Dynkin labels of the highest weight are all non-negative, $a_a(W) \geq 0$, in analogy to the highest root. 

Instead of indicating the highest weights we often use the standard physics notation where we indicate irreducible representations by their dimension. Since we only encounter adjoints of $\text{A},\text{D},\text{E}$, (anti-symmetrizations of) fundamental irreducible representations of $\text{A}_N$, vectorial and spinorial irreducible representations of $\text{D}_N$, and the lowest-dimensional irreducible representations of $\text{E}_N$, this notation is unique up to complex conjugate irreducible representations, which are distinguished by adding a bar over one of them. If needed, we add the subscript $+/-$ to distinguish the spinor and cospinor representations of $\text{D}_N$ algebras.\footnote{For SO(8) this then also uniquely distinguishes the vector $\rep{8}$ from both spinor representations $\rep{8}_\pm$.} 

The highest weight of the $k$-fold anti-symmetrized fundamental representation of A$_N$ has Dynkin labels $a_a=\delta_{a,k}$, corresponding to a Young tableau with a single column and $k$ rows. Starting from this highest weight, all other weights of the irreducible representation can be constructed via the highest weight procedure, where one consecutively subtracts simple roots. It is also simpler to do this using Dynkin labels; if $a_a=l$ this means that one can descend $l$ times with the $a^\text{th}$ simple root. When using Dynkin labels, this amounts to subtracting the $a^\text{th}$ row of the Cartan matrix. The procedure ends if no positive Dynkin labels are left after descending with the simple roots as often as possible. The highest weight of the adjoint representation of A$_N$ has Dynkin labels $a_{1}=a_{N}=1$ and the rest of the $a_a=0$.

For D$_N$, the highest weight of the vector irreducible representation has Dynkin labels $a_a=\delta_{a,1}$, the one of the adjoint, which is (contained in) the two-fold antisymmetrized vector irreducible representation, has Dynkin labels $a_a=\delta_{a,2}$, and the two highest weights of the spinor representations have Dynkin labels $a_a=\delta_{a,N-1}$ and $a_a=\delta_{a,N}$, respectively. 

For E$_6$ the highest weight\footnote{For $\E6$ and $\E7$ we use a similar labeling as for $\E8$, i.e.\ we start counting at the long side of the chain of nodes and assign the highest number to the single node that stands out.} of the lowest-dimensional irreducible representation $27$ has $a_a=\delta_{a,1}$ and the adjoint $78$ has $a_a=\delta_{a,6}$. For E$_7$ the highest weight of the lowest-dimensional irreducible representation $56$ has $a_a=\delta_{a,1}$ and the adjoint $133$ has $a_a=\delta_{a,6}$. For E$_8$, the adjoint $248$ is the lowest-dimensional irreducible representation with highest weight Dynkin label $a_a=\delta_{a,1}$.

\subsection{Lattices} 
\label{sc:Lattices} 

In order to have a universal description for all heterotic string theories, it is convenient to choose a Cartan subalgebra such that the roots can either be written as SO($2N$)-adjoint roots $\big(\,\undr{\pm 1^2,0^{N-2}}\,\big)$ or SO($2N$)-spinor weights $\big(\,\undr{-\sfrac 12^{k}, \sfrac12^{N-k}}\,\big)$, $k \in \Natr$. Here, a superscript signifies that the corresponding entry is repeated this power number of times. Furthermore, an underscore means that all possible permutations of the underlined entries are taken (with all possible signs for the non-zero entries indicated by $\pm$). These roots and weights span various lattices listed in Table~\ref{tb:Lattices}. 

\begin{table}
\[
\arry{|c||c|c|c|c|}{
\hline 
\text{\bf Lattice} & \mathbf{R}_D & \mathbf{V}_D & \mathbf{S}_D & \mathbf{S}_D 
\\ \hline\hline 
\text{\bf Entries} & ~~~\text{integral}~~~ & ~~~\text{integral}~~~ & \text{half-integral} & \text{half-integral} 
\\
\text{\bf Sum} & \text{even} & \text{odd} & \text{even} & \text{odd}  
\\
\text{\bf Generators} & \big(\,\undr{\pm 1^2, 0^{14}}\,\big) &  \big(\,\undr{\pm 1, 0^{15}}\,\big) & 
\big(\,\undr{-\sfrac12^{2k}, \sfrac12^{16-2k}}\,\big) & \big(\,\undr{-\sfrac12^{2k+1}, \sfrac12^{15-2k}}\,\big) 
\\ \hline 
}
\]
\caption{\label{tb:Lattices}
The four different lattices that appear in heterotic string theories are listed. An underscore denotes all permutations of the underlined entries.}
\end{table}

\subsection{Branching of Lie algebras}
\label{sc:Branchings} 

In this work we make use of two branchings of Lie groups: Levi-type and extended branchings. Below we briefly describe both of them.

\subsubsection*{Extended branchings}

In an extended branching one obtains a maximal semi-simple subgroup of the group $\mathcal{G}$.  This type of branching is obtained by deleting a single node from the extended Dynkin diagram of $\mathcal{G}$. By deleting the $k^\text{th}$ node from the extended Dynkin diagram of E$_8$, one obtains in this way a chain of subgroups E$_8\supset\text{E}_{8-k}\times \text{A}_k$ where we identify by abuse of notation E$_5=\text{D}_5$, E$_4=\text{A}_4$, E$_3=\text{A}_2\times \text{A}_1$, E$_2=\text{A}_1\times \text{A}_1$, E$_1=\text{A}_1$, E$_0=\mathbbm{1}$. For D$_N$, deleting the $k^\text{th}$ node ($k>2$) from the extended Dynkin diagram gives the chain $\text{D}_N\supset \text{D}_k\times \text{D}_{N-k}$ with $\text{D}_3\simeq \text{A}_3$. Deleting the second node gives $\text{D}_N\supset \text{A}_1\times \text{A}_1\times \text{D}_{N-2}$ while deleting nodes zero or one gives the original Dynkin diagram.

In order to obtain the new root system of the Lie algebras after branching one can use the projection matrix which specifies how the simple roots of the original gauge algebra are mapped onto those of the subalgebras. For the extended branching, one replaces the deleted root $\alpha_k$ by the extended root such that the projection matrix reads
\begin{align} \label{ProjectionMatrix} 
 P_{ab}=\left\{\begin{array}{ll}
 \delta_{ab} & a\neq k~, \\ 
 c_b \,\delta_{ak} & a=k~.
 \end{array}\right.
\end{align}
The irreducible representations into which a given representation of $\mathcal{G}$ branches can be obtained by applying the projection matrix to all weights of the representation in question. There also exists a Mathematica package to preform such manipulations~\cite{Feger:2012bs}.

Let us consider the breaking of $\text{E}_8$ to $\text{SU(5)}\times\text{SU(5)}$ as a concrete example. As can be seen from Figure~\ref{fig:ADEDynkin}, the extended branching corresponds to removing the fourth node from the extended Dynkin diagram. The simple roots of the first $\text{A}_4$ are then $(\alpha_0,\alpha_1,\alpha_2,\alpha_3)$ and those of the second $\text{A}_4$ are $(\alpha_8,\alpha_5,\alpha_6,\alpha_7)$. (Here we have already reordered the roots such that they match the standard numbering convention.) Using the Dynkin multiplicities $c_a$, which are also given in Figure~\ref{fig:ADEDynkin}, the branching of $\text{E}_8 \supset \text{SU(5)} \times \text{SU(5)}$ can be described by the projection matrix 
\begin{equation}
\label{eq:SU5ProjectionMatrix-Ext}
 P_{\text{E}_8\supset \text{A}_4\times \text{A}_4} =
 \left(
 \scalebox{.6}{$
 \begin{matrix}
  2&3&4&5&6&4&2&3\\
  1&0&0&0&0&0&0&0\\
  0&1&0&0&0&0&0&0\\
  0&0&1&0&0&0&0&0\\
  0&0&0&0&0&0&0&1\\
  0&0&0&0&1&0&0&0\\
  0&0&0&0&0&1&0&0\\
  0&0&0&0&0&0&1&0
 \end{matrix}$}
 \right)~.
\end{equation}
In order to compute the irreducible representations into which the adjoint of E$_8$ branches one can apply the projection matrix to all 248 roots of E$_8$ and identify the irreducible representations of the branched gauge groups to which each root belongs after branching.

\subsubsection*{Levi-type branchings}

In a Levi-type branching the group $\mathcal{G}$ is branched to a subgroup that contains one or more U(1) factors. This branching into maximal non-semisimple subgroups is obtained by deleting $r$ nodes from the ordinary Dynkin diagram of $\mathcal{G}$. Each deleted node turns into a U(1) factor; the remaining roots describe the semi-simple part. 

The choice of U(1) basis for $r$ U(1) factors is rather arbitrary. A specific choice for the U(1) charges of a representation with corresponding weight $w$ can be obtained by taking the inner product of the $a^\text{th}$ row of the inverse Cartan matrix of $\mathcal{G}$ with $w$,
\begin{align} \label{CanonicalCharges} 
 Q_a = (A^{-1})_{ab} w_b\,, 
\end{align}
for the U(1) factors labeled by $a=1,\ldots, r$.

We can encode this information in an extension of the projection matrix: Since for Levi-breaking the semi-simple part of the unbroken gauge group does not have maximal rank, $r$ rows of the projection matrix are undetermined. We account for this in the projection matrix by substituting the $a^\text{th}$ row by the $a^\text{th}$ row of the inverse Cartan matrix. To clearly indicate the different functions of these different rows we separate them by a line.

Let us again consider an $\text{E}_8$ branching example, this case to  $\text{SU(5)}\times \text{U(1)}^4$. The Levi-type branching corresponds to removing nodes $1,2,3,4$ from the ordinary Dynkin diagram. The four U(1) charges of the Cartan generators that have been removed are given in terms of the first four rows of the inverse Cartan matrix. Upon permutations of the rows, we obtain the projection matrix
\begin{equation}
\label{eq:SU5ProjectionMatrix-Levi}
 P_{\text{E}_8\supset \text{A}_4\times \text{U}(1)^4} =
 \left(
 \scalebox{.6}{$
 \begin{matrix}
  0&0&0&0&0&0&0&1\\
  0&0&0&0&1&0&0&0\\
  0&0&0&0&0&1&0&0\\
  0&0&0&0&0&0&1&0
  \\ \hline 
  8&16&24&20&15&10&5&12\\
  6&12&18&15&12&8&4&9\\
  4&8&12&10&8&6&3&6\\
  2&4&6&5&4&3&2&3
 \end{matrix}$}
 \right)~. 
\end{equation}

\def\theequation{\thesection.\arabic{equation}} 
\setcounter{equation}{0}
\section[A bundle vector representation of \texorpdfstring{S(U(1)$^{5}$}{S(U(1)\^{}5}) bundles]{A bundle vector representation of \texorpdfstring{S(U(1)$^{\boldsymbol{5}}$}{S(U(1)\^{}5}) bundles}
\label{sc:IdUbundles}

Bundles of the type $\text{S}(\text{U}(1)^5) \subset \text{E}_8$ have been frequently considered in the literature, see e.g.~\cite{Anderson:2011ns,Anderson:2012yf,Anderson:2013xka}. In~\cite{Nibbelink:2015ixa,Blaszczyk:2015zta} it was suggested that this description can be reformulated in terms of line bundle vectors $V_i$: In order to obtain an unbroken  SU(5) group one may choose the line bundle vectors 
\equ{ \label{SU5inE8emb} 
V_i = (a_i^5, b_i, c_i, d_i)~, 
}
assuming that the parameters $a_i\neq0, b_i, c_i, d_i$ are sufficiently generic. This parameterization can be related to the vectors $k_i$ by comparing the charges of the states that appear in the branching 
\equ{ 
\mathbf{248} \ra 
(\rep{24},\rep{1}) + (\rep{1},\rep{24}) + (\rep{10}, \rep{5}) + (\crep{10}, \crep{5})  
+ (\rep{5}, \crep{10}) + (\crep{5}, \rep{10}) 
}
under $\text{E}_8 \supset \text{SU}(5)\times \text{SU}(5)$. In particular, for the $\rep{10}$-plets this leads to the charge table: 
\equ{ 
\arry{|c|c|c|}{ 
\hline 
\mathbf{10}\text{\bf -plets} & V_i\text{\bf -charges} & k_i\text{\bf -charges} 
\\\hline\hline 
(\underline{\sm \sfrac 12^3,\sfrac 12^2}, \sm \sfrac 12, ~\,\sfrac 12, ~\,\sfrac 12) &  
- \sfrac {a_i}2 - \sfrac {b_i}2 + \sfrac {c_i}2 + \sfrac {d_i}2  & k^{1}_i 
\\ 
(\underline{\sm \sfrac 12^3,\sfrac 12^2}, ~\,\sfrac 12, \sm\sfrac 12, ~\,\sfrac 12) &  
- \sfrac {a_i}2 + \sfrac {b_i}2 - \sfrac {c_i}2 + \sfrac {d_i}2  & k^{2}_i 
\\ 
(\underline{\sm \sfrac 12^3,\sfrac 12^2},  ~\,\sfrac 12, ~\,\sfrac 12, \sm\sfrac 12) &  
- \sfrac {a_i}2 + \sfrac {b_i}2 + \sfrac {c_i}2 - \sfrac {d_i}2  & k^{3}_i 
\\ 
(\underline{\sm \sfrac 12^3,\sfrac 12^2}, \sm \sfrac 12, \sm\sfrac 12, \sm\sfrac 12) &  
- \sfrac {a_i}2 - \sfrac {b_i}2 - \sfrac {c_i}2 - \sfrac {d_i}2  & k^{4}_i 
\\ 
(\underline{1^2,0^3}, 0^3)
& 2\, a_i & k^{5}_i
\\\hline 
}\nonumber
}
This table expresses the line bundle vector parameters in terms of the quantities $k_i^{\tilde{a}}$ which is easily inverted: 
\equ{
\arry{lcl}{
a_i = - \sfrac 12\Big( ~\, k_i^{1} +k_i^{2} +k_i^{3} +k_i^{4} \Big)~, 
&\qquad &
b_i = - \sfrac 12\Big( ~\, k_i^{1} - k_i^{2} -k_i^{3} +k_i^{4} \Big)~, 
\\[1ex] 
c_i = - \sfrac 12\Big( \sm k_i^{1} +k_i^{2} -k_i^{3} +k_i^{4} \Big)~, 
&\qquad &
d_i = - \sfrac 12\Big(\sm k_i^{1} - k_i^{2} +k_i^{3} +k_i^{4} \Big)~. 
}
}

\end{appendices}

\providecommand{\href}[2]{#2}\begingroup\raggedright\endgroup

\end{document}